\documentclass[reqno]{amsart}

\usepackage[OT1]{fontenc}
\usepackage[utf8]{inputenc}

\usepackage{mathtools}
\usepackage{subfigure,graphicx}
\usepackage{sidecap}
\usepackage{color}
\usepackage{natbib}
\usepackage{hyperref}
\hypersetup{colorlinks,urlcolor=blue,citecolor=blue}

\usepackage{threeparttable} 

\newtheorem{theorem}{Theorem}[section]
\newtheorem{proposition}[theorem]{Proposition}
\newtheorem{lemma}[theorem]{Lemma}
\theoremstyle{definition}

\theoremstyle{remark}
\newtheorem{remark}[theorem]{Remark}

\numberwithin{equation}{section}

\DeclareMathOperator{\AIC}{AIC}

\DeclareMathOperator{\ARE}{ARE}
\DeclareMathOperator{\argmax}{arg\,max}
\DeclareMathOperator{\bias}{bias}
\DeclareMathOperator{\cov}{Cov}

\def\dev{\mathrm{DEV}}
\DeclareMathOperator{\diag}{diag}
\DeclareMathOperator{\E}{E}
\DeclareMathOperator{\IF}{IF}
\DeclareMathOperator{\IQR}{IQR}

\def\ML{\mathrm{ML}}
\def\opt{\mathrm{opt}}

\def\qtl{\mathrm{QTL}}
\def\Rset{\mathbb{R}}
\DeclareMathOperator{\rd}{d}
\DeclareMathOperator{\rk}{rank}

\DeclareMathOperator{\sign}{sign}
\DeclareMathOperator{\tr}{tr}
\DeclareMathOperator{\var}{var}

\newcommand{\bm}[1]{\mbox{\boldmath $#1$}} 
\newcommand{\what}[1]{\widehat{#1}}

\makeatletter
\def\tagform@#1{\maketag@@@{\color{blue}(\ignorespaces#1\unskip\@@italiccorr)}}
\makeatother

\DeclareGraphicsExtensions{.pdf,.jpg,.png}

\counterwithin{table}{section}
\counterwithin{figure}{section}

\begin{document}

\title[A robust approach for generalized linear models]{A robust approach for generalized linear models based on maximum L$\bm{q}$-likelihood procedure}

\author{Felipe Osorio}
\address{Departamento de Matem\'atica, Universidad T\'ecnica Federico Santa Mar\'ia, Chile}
\curraddr{Avenida Espa\~na 1680, Valpara\'iso, Chile}
\email{\href{mailto:felipe.osorios@usm.cl}{felipe.osorios@usm.cl}}
\thanks{Corresponding author. E-mail address: \href{mailto:felipe.osorios@usm.cl}{\texttt{felipe.osorios@usm.cl}} (F. Osorio)}

\author{Manuel Galea}
\address{Departamento de Estad\'istica, Pontificia Universidad Cat\'olica de Chile} 
\curraddr{Avenida Vicu\~na Mackena 4860, Santiago, Chile}
\email{\href{mailto:mgalea@mat.puc.cl}{mgalea@mat.puc.cl}}

\author{Patricia Gim\'enez}
\address{Departamento de Matem\'atica, Universidad Nacional de Mar del Plata, Argentina}
\curraddr{Funes 3350, CP 7600, Mar del Plata, Argentina.}
\email{\href{mailto:pcgimene@mdp.edu.ar}{pcgimene@mdp.edu.ar}}

\maketitle

\begin{abstract}
  In this paper we propose a procedure for robust estimation in the context of generalized
  linear models based on the maximum L$q$-likelihood method. Alongside this, an estimation 
  algorithm that represents a natural extension of the usual iteratively weighted least 
  squares method in generalized linear models is presented. It is through the discussion of 
  the asymptotic distribution of the proposed estimator and a set of statistics for testing 
  linear hypothesis that it is possible to define standardized residuals using the mean-shift 
  outlier model. In addition, robust versions of deviance function and the Akaike information 
  criterion are defined with the aim of providing tools for model selection. Finally, the 
  performance of the proposed methodology is illustrated through a simulation study and 
  analysis of a real dataset.

\end{abstract}

\section{Introduction}\label{sec:Intro}

In this work we consider the robust estimation of the regression coefficients in generalized
linear models \citep{Nelder:1972}. This model is defined by asumming that $y_1,\dots,y_n$
are independent random variables, each with density function
\begin{equation}\label{eq:density}
  f(y_i;\theta_i,\phi) = \exp[\phi\{y_i\theta_i - b(\theta_i)\} + c(y_i,\phi)],
  \qquad i = 1,\dots,n,
\end{equation}
where $\phi > 0$ represents a scale parameter, and $b(\cdot)$ and $c(\cdot,\cdot)$
are known functions. It is also assumed that the systematic part of the model is given
by a linear predictor $\eta_i=\bm{x}_i^\top\bm{\beta}$, which is related to the natural 
parameter
\begin{equation}\label{eq:linear}
  \theta_i = k(\eta_i), \qquad i=1,\dots,n,
\end{equation}
by a one-to-one differentiable function $k(\cdot)$, this is called $\theta$-link 
to distinguish it from the conventional link that relates the covariates to the mean of 
$y_i$. The $\theta$-link function has been used by \cite{Thomas:1989, Thomas:1990} to 
perform local influence analyses in generalized linear models. The aforementioned allows 
us to write $\bm{\eta} = \bm{X\beta}$, where $\bm{X}$ is a $n\times p$ design matrix with full 
column rank and $\bm{\beta} = (\beta_1,\dots,\beta_p)^\top$ represents the unknown regression 
coefficients.

A number of authors have considered procedures for robust estimation of regression
coefficients in generalized linear models (GLM), with particular emphasis on logistic regression
\citep[see][]{Pregibon:1982, Stefanski:1986, Kunsch:1989, Morgenthaler:1992}, whereas
\cite{Cantoni:2001} developed an approach to obtain robust estimators in the framework
of quasi-likelihood based on the class of $M$-estimators of Mallows' type proposed by
\cite{Preisser:1999}. More recently, \cite{Ghosh:2016} considered an alternative
perspective for robust estimation in generalized linear models based on the minimization
of the density power divergence introduced by \cite{Basu:1998}.

To address the robust estimation in the framework of generalized linear models, we will
briefly describe the estimation procedure introduced by \cite{Ferrari:2010} \citep[see 
also][]{Ferrari:2009} who proposed the minimization of a $q$-divergence and showed it to 
be equivalent to the minimization of an empirical version of Tsallis-Havrda-Charv\'at
entropy \citep{Havrda:1967,Tsallis:1988} also known as $q$-entropy. An interesting feature
of this procedure is that the efficiency of the method depends on a tuning parameter $q$,
known as distortion parameter. It should be stressed that the methodology proposed as part 
of this work offers a parametric alternative to perform robust estimation in generalized 
linear models which allows, for example, to obtain robust estimators for logistic regression, 
an extremely relevant input to carry out binary classification.

\cite{Ferrari:2010} introduced a procedure to obtain robust estimators, known as maximum
L$q$-likelihood estimation, which is defined as
\[
  \what{\bm{\beta}}{}_q^* = \underset{\beta\in\Rset^p}{\argmax}\ L_q(\bm{\beta}), \qquad q > 0,
\]
where
\begin{equation}\label{eq:objective}
  L_q(\bm{\beta}) = \sum_{i=1}^n l_q(f(y_i;k(\bm{x}_i^\top\bm{\beta}),\phi)),
\end{equation}
with $l_q(u)$ denoting the deformed logarithm of order $q$, whose definition is given by:
\[
  l_q(u) = \begin{cases}
    (u^{1-q} - 1) / (1 - q), & \quad q\neq 1, \\
    \log(u), & \quad q=1.
  \end{cases}
\]
The function $l_q(\cdot)$ is known in statistics as the Box-Cox transformation, 
The notation for the objective function defined in \eqref{eq:objective} emphasizes the fact that 
we are considering $\phi$ as a nuisance parameter. Note that this parameter is in fact known
for many models of the exponential family.

It should be stressed that, in general, the maximum L$q$-likelihood estimator, $\what{\bm{\beta}}{}_q^*$ 
is not Fisher consistent for the true value of parameter $\bm{\beta}_0$. An appropriate transformation 
$\tau_q$ in order to obtain a Fisher consistent version of the maximum L$q$-likelihood estimator 
is considered in next section. Usually, the maximum L$q$-likelihood estimator $\what{\bm{\beta}}{}_q^*$ 
is obtained by solving the system of equations $\bm{\Psi}_n(\bm{\beta}) = \bm{0}$, which assume the form
\begin{align*}
  \bm{\Psi}_n(\bm{\beta}) & = \sum_{i=1}^n\frac{\partial}{\partial\bm{\beta}}\,l_q(f(y_i;k(\bm{x}_i^\top\bm{\beta}),\phi)) \\
  & = \sum_{i=1}^n\,\{f(y_i;k(\bm{x}_i^\top\bm{\beta}),\phi)\}^{1-q}\frac{\partial}{\partial\bm{\beta}}
  \log f(y_i;k(\bm{x}_i^\top\bm{\beta}),\phi).
\end{align*}
Therefore, $\bm{\Psi}_n(\bm{\beta})$ can be written as an estimation function associated to the
solution of a weighted likelihood, as follows
\begin{equation}\label{eq:weighted-psi}
  \bm{\Psi}_n(\bm{\beta}) = \sum_{i=1}^n \bm{\Psi}_i(\bm{\beta}) = \sum_{i=1}^n U_i(\bm{\beta})
  \bm{s}_i(\bm{\beta}),
\end{equation}
with $U_i(\bm{\beta}) = \{f(y_i;k(\bm{x}_i^\top\bm{\beta}),\phi)\}^{1-q}$ being a weighting based 
on a power of the density function and $\bm{s}_i(\bm{\beta}) = \partial\log f(y_i;k(\bm{x}_i^\top\bm{\beta}),
\phi)/\partial\bm{\beta}$, denoting the score function associated with the $i$th observation
$(i=1,\dots,n$). Interestingly, for $q\neq 1$, the estimation procedure attributes
weights $U_i(\bm{\beta})$ to each of the observations depending on their probability of
occurrence. Thus, those observations that disagree with the postulated model will receive
a small weighting in the estimation procedure as long as $q < 1$, while if $q > 1$
the importance of those observations whose density is close to zero will be accentuated.
It should be noted that the maximum L$q$-likelihood estimation procedure corresponds
to a generalization of the maximum likelihood method. Indeed, it is easy to notice
that for $q\to 1$, we obtain that $l_q(u)\to\log(u)$, and hence $\what{\bm{\beta}}{}_q^*$ 
converges to the maximum likelihood estimator of $\bm{\beta}$.

When $q$ is fixed, the maximum L$q$-likelihood estimator $\what{\bm{\beta}}{}_q^*$ belongs to 
the class of $M$-estimators and we can use the results available in \cite{Hampel:1986} to
study the statistical properties of this kind of estimators. Since such estimators are
defined by optimizing the objective function given in \eqref{eq:objective}, we can also
use the very general setting of extremum estimation \citep{Gourieroux:1995} to obtain
their asymptotic properties. A further alternative is to consider $\what{\bm{\beta}}{}_q^*$ 
as a solution of an inference function and invoke results associated with asymptotic normality
that have been developed for that class of estimators \citep[see, for instance][]{Yuan:1998}.
In particular, \cite{Ferrari:2010} derived the asymptotic properties of the maximum
L$q$-likelihood estimator considering distributions belonging to the exponential family.
An interesting result of that work is that the asymptotic behavior of this estimator
is characterized by the distortion parameter $q$.

Although there are relatively few applications of this methodology in the statistical
literature, we can highlight that this procedure has been extended to manipulate models
for incomplete data \citep{Qin:2013}, to carry out hypothesis tests using a generalized
version of the likelihood ratio statistic \citep{Huang:2013, Qin:2017} as well as to
address the robust estimation in measurement error models \citep{Gimenez:2022a, Gimenez:2022b}
and beta regression \citep{Ribeiro:2023}.

The remainder of the paper unfolds as follows. We begin by describing the estimation procedure 
in GLM based on maximum L$q$-likelihood, discussing its asymptotic properties and some methods 
for the selection of the distortion parameter. Statistics for testing linear hypotheses, model 
selection methods and the definition of standardized residuals are discussed in Section 
\ref{sec:inference}, whereas Section \ref{sec:exp} illustrates the good performance of our 
proposal by analyzing a real data set as well as a Monte Carlo simulation experiment. Appendices 
include proofs of all the theoretical results.

\section{Robust estimation in GLM based on maximum L$q$-likelihood}

\subsection{Estimation procedure and asymptotic properties}

The maximum L$q$-likelihood estimator for $\bm{\beta}$ in the class of generalized linear
models is a solution of the estimating equation:
\begin{equation}\label{eq:psi-GLM}
  \bm{\Psi}_n(\bm{\beta}) = \phi\sum_{i=1}^n U_i(\bm{\beta})W_i^{1/2}\frac{(y_i - \mu_i)}{\sqrt{V_i}}\bm{x}_i,
\end{equation}
with
\begin{equation}\label{eq:robweights}
  U_i(\bm{\beta}) = \exp\{(1-q)\phi[y_ik(\bm{x}_i^\top\bm{\beta}) - b(k(\bm{x}_i^\top\bm{\beta}))] + c_q(y_i,\phi)\},
\end{equation}
corresponds to the weight arising from the estimation function defined in \eqref{eq:weighted-psi},
with $c_q(y_i,\phi) = (1-q)c(y_i,\phi)$, while $\mu_i = \E(y_i) = \dot{b}(\theta_i)$, $\var(y_i)
= \phi^{-1}V_i$, with $V_i = \ddot{b}(\theta_i)$ being the variance function, and $W_i = V_i
\{\dot{k}(\eta_i)\}^2$, we shall use dots over functions to denote derivatives, $\dot{k}(u) =
{\rd}k/{\rd}u$ and $\ddot{k}(u) = {\rd}^2k/{\rd}u^2$. In this paper, we consider $q < 1$ in
which case the weights $U_i(\bm{\beta})$ $(i=1\dots,n)$ provide a mechanism for downweighting
outlying observations. The individual score function adopts the form $\bm{s}_i(\bm{\beta}) = \phi
W_i^{1/2}r_i\bm{x}_i$, where $r_i = (y_i - \mu_i)/\sqrt{V_i}$ denotes the Pearson residual. It
is straightforward to notice that the estimation function for $\bm{\beta}$ can be written as:
\[
  \bm{\Psi}_n(\bm{\beta}) = \phi \bm{X}^\top \bm{W}^{1/2}\bm{UV}^{-1/2}(\bm{Y} - \bm{\mu}),
\]
where $\bm{U} = \diag(U_1,\dots,U_n)$, with $U_i = U_i(\bm{\beta})$ is given in \eqref{eq:robweights},
while $\bm{W} = \diag(W_1,\dots,W_n)$, $\bm{V} = \diag(V_1,\dots,V_n)$, $\bm{Y} = (y_1,\dots,y_n)^\top$
and $\bm{\mu} = (\mu_1,\dots,\mu_n)^\top$. 

In Appendix \ref{app:Fisher} we show that the estimating function in \eqref{eq:psi-GLM} is unbiased 
for the surrogate parameter $\bm{\beta}_*$, that is $\E_0\{\bm{\Psi}_n(\bm{\beta}_*)\} = \bm{0}$, where 
$\E_0(\cdot)$ denotes expectation with respect to the true distribution $f(\cdot;\theta_0,\phi)$. Thereby, 
this allows introducing a suitable calibration function, say $\tau_q(\cdot)$, such that $\what{\bm{\beta}}_q 
= \tau_q(\what{\bm{\beta}}{}_q^*)$ which is applied to rescale the parameter estimates in order to achieve 
Fisher-consistency for $\bm{\beta}_0$.

When partial derivatives and expectations exist, we define the matrices
\begin{align}
  \bm{A}_n(\bm{\beta}) & = \E_0\{\bm{\Psi}_n(\bm{\beta})\bm{\Psi}_n^\top(\bm{\beta})\} 
  = \frac{\phi}{2-q}\bm{X}^\top\bm{WJX}, \label{eq:variability} \\
  \bm{B}_n(\bm{\beta}) & = \E_0\Big\{-\frac{\partial\bm{\Psi}_n(\bm{\beta})}{\partial\bm{\beta}^\top}\Big\}
  = \phi \bm{X}^\top \bm{WJGKX}.
  \label{eq:sensitivity}
\end{align}
where $\bm{J} = \diag(J_1,\dots,J_n)$ whose diagonal elements are given by $J_i = J_q^{-\phi}(\theta_i)$,
for $i=1,\dots,n$, with $J_q(\theta) = \exp\{-qb(\theta) + b(q\theta)\}$, $q\theta\in\Theta$ was
defined by \cite{Menendez:2000}, whereas $\bm{G} = \diag(\dot{g}(\theta_1^*),\dots,\dot{g}(\theta_n^*))$
and $\bm{K} = \diag(\dot{k}(\eta_1),\dots,\dot{k}(\eta_n))$ with ${\rd}\eta_*/{\rd}\eta = (1/q)\dot{g}(\theta_*)
\dot{k}(\eta)$ and $g(\theta_*) = k^{-1}(\theta_*)$ corresponding to the inverse function of the 
$\theta$-link evaluated on the surrogate parameter. We propose to use the Newton-scoring algorithm 
\citep{Jorgensen:2004} to address the parameter estimation associated with the estimating equation 
$\bm{\Psi}_n(\bm{\beta}) = \bm{0}$, which assumes the form
\[
  \bm{\beta}^{(t+1)} = \bm{\beta}^{(t)} + \bm{B}_n^{-1}(\bm{\beta}^{(t)})\bm{\Psi}_n(\bm{\beta}^{(t)}), \qquad t=0,1,\dots,
\]
Therefore, the estimation procedure
adopts an iteratively reweighted least squares structure, as
\begin{equation}\label{eq:IWLS}
  \bm{\beta}^{(t+1)} = (\bm{X}^\top \bm{W}^{(t)}\bm{J}^{(t)}\bm{G}^{(t)}\bm{K}^{(t)}\bm{X})^{-1}
  \bm{X}^\top \bm{W}^{(t)}\bm{J}^{(t)}\bm{G}^{(t)}\bm{K}^{(t)}\bm{Z}_q^{(t)},
\end{equation}
where $\bm{Z}_q = \bm{\eta} + \bm{J}^{-1}\bm{G}^{-1}\bm{K}^{-1}\bm{W}^{-1/2}\bm{UV}^{-1/2}(\bm{Y} - \bm{\mu})$ 
denotes the working response which must be evaluated at $\bm{\beta} = \bm{\beta}^{(t)}$. Let $\what{\bm{\beta}}{}_q^*$ 
be the value obtained at the convergence of the algorithm in \eqref{eq:IWLS}. Thus, in order
to obtain a corrected version of the maximum L$q$-likelihood estimator, we must take 
\begin{equation}\label{eq:calibration}
  \what{\bm{\eta}}_q = k^{-1}(qk(\what{\bm{\eta}}{}_q^*)), 
  \quad \textrm{with} \quad \what{\bm{\eta}}{}_q^* = \bm{X}\what{\bm{\beta}}{}_q^*.
\end{equation}
Evidently, for the canonical $\theta$-link, $k(u)=u$, it is sufficient to consider $\what{\bm{\eta}}_q 
= q\what{\bm{\eta}}{}_q^*$, or equivalently $\what{\bm{\beta}}_q = q\what{\bm{\beta}}{}_q^*$. In addition, 
for this case we have $\bm{GK} = \bm{I}_n$, with $\bm{I}_n$ being the identity matrix of order $n$, which 
leads to a considerably simpler version of the algorithm proposed in \eqref{eq:IWLS}. \cite{Qin:2017} suggested 
another alternative to achieve Fisher-consistency which is based on re-centering the estimation equation, 
considering a bias-correction term, and who termed it as the bias-corrected maximum L$q$-likelihood 
estimator. We emphasize the simplicity of the calibration function in \eqref{eq:calibration}, which 
introduces a reparameterization that allows us to characterize the asymptotic distribution of $\what{\bm{\beta}}_q$.

\begin{remark}
  For $q = 1$, it is easy to notice that in the formula given in \eqref{eq:IWLS}, we obtain 
  $\bm{J} = \bm{I}_n$ and $\bm{U} = \bm{I}_n$, thereby leading to the Fisher-scoring method for 
  maximum likelihood estimation in generalized linear models \citep[see, for instance,][]{Green:1984}.
\end{remark}

The following result describes the asymptotic normality of the $\what{\bm{\beta}}_q$ estimator
in the context of generalized linear models.

\begin{proposition}
  For the generalized linear model defined in \eqref{eq:density} and \eqref{eq:linear},
  and under assumptions of Property 24.16 of \cite{Gourieroux:1995}, it follows that
  \[
    (\bm{B}_n^{-1}\bm{A}_n\bm{B}_n^{-1})^{-1/2}\sqrt{n}(\what{\bm{\beta}}_q - \bm{\beta}_0) 
    \stackrel{D}{\longrightarrow} \mathsf{N}_p(\bm{0},\bm{I}_p),
  \]
  where $\bm{A}_n = \bm{A}_n(\bm{\beta}_0)$ and $\bm{B}_n = \bm{B}_n(\bm{\beta}_0)$. That is, 
  $\sqrt{n}(\what{\bm{\beta}}_q - \bm{\beta}_0)$ asymptotically follows a $p$-variate normal 
  distribution with mean vector $\bm{0}$ and covariance matrix,
  \[
    \bm{B}_n^{-1}\bm{A}_n\bm{B}_n^{-1} = \frac{\,\phi^{-1}}{2 - q}(\bm{X}^\top \bm{WJGKX})^{-1}
    \bm{X}^\top \bm{WJX}(\bm{X}^\top \bm{WJGKX})^{-1}.
  \]
\end{proposition}

\begin{remark}
  It is straightforward to notice that for the canonical $\theta$-link $k(u) = u$ the asymptotic
  covariance matrix of the $\what{\bm{\beta}}_q$ estimator adopts a rather simple form, namely:
  \[
    \bm{B}_n^{-1}\bm{A}_n\bm{B}_n^{-1} = \frac{\,\phi^{-1}}{2 - q}(\bm{X}^\top \bm{WJX})^{-1}.
  \]
\end{remark}

\subsection{Influence function and selection of distortion parameter $\bm{q}$}\label{sec:IF}

From the general theory of robustness, we can characterize a slight misspecification
of the proposed model by the influence function of a statistical functional $\bm{T}(F)$,
which is defined as \citep[see][Sec.\,4.2]{Hampel:1986},
\[
  \IF(Y;\bm{T}) = \lim_{\epsilon\to 0}\frac{\bm{T}(F_\epsilon) - \bm{T}(F)}{\epsilon}
  = \frac{\partial \bm{T}(F_\epsilon)}{\partial\epsilon}\Big|_{\epsilon = 0},
\]
where $F_\epsilon = (1-\epsilon)F + \epsilon\delta_Y$, with $\delta_Y$ a point mass
distribution in $Y$. That is, $\IF(Y;\bm{T})$ reflects the effect on $\bm{T}$ of a contamination
by adding an observation in $Y$. It should be noted that the maximum L$q$-likelihood
estimation procedure for $q$ fixed corresponds to an $M$-estimation method, where $q$
acts as a tuning constant. It is well known that the influence functions of the estimators
obtained by the maximum likelihood and L$q$-likelihood methods adopt the form
\[
  \IF(Y;\what{\bm{\beta}}_\ML) = \bm{\mathcal{F}}_n^{-1}\bm{s}(\bm{\beta}), \qquad
  \IF_q(Y;\what{\bm{\beta}}_q) = \bm{B}_n^{-1}\{f(y;k(\bm{x}^\top\bm{\beta}),\phi)\}^{1-q}\bm{s}(\bm{\beta}),
\]
respectively, where $\bm{\mathcal{F}}_n = \E\{-\partial \bm{s}_n(\bm{\beta})/\partial\bm{\beta}^\top\}
= \phi^{-1}\bm{X}^\top\bm{WX}$ is the Fisher information matrix, with $\bm{s}_n(\bm{\beta}) = \sum_{i=1}^n
\bm{s}_i(\bm{\beta})$ the score function, whereas $\{f(y;k(\bm{x}^\top\bm{\beta}),\phi)\}^{1-q}$ corresponds
to the weighting defined in \eqref{eq:robweights} and $\bm{s}(\bm{\beta}) = \phi W^{1/2}(y
- \mu)\bm{x}/\sqrt{V}$, with $\bm{x} = (x_1,\dots,x_p)^\top$. It is possible to appreciate that
when $q < 1$ the weights $U_i(\bm{\beta})$ provide a mechanism to limit the influence of
extreme observations. This allows downweighting those observations that are strongly
discrepant from the proposed model. The properties of robust procedures based on the
weighting of observations using the working statistical model have been discussed,
for example in \cite{Windham:1995} and \cite{Basu:1998}, who highlighted the connection
between these procedures with robust methods \citep[see also,][]{Ferrari:2010}. Indeed,
we can write the asymptotic covariance of $\what{\bm{\beta}}_q$ equivalently as,
\[
  \cov(\what{\bm{\beta}}_q) = \E_0\{\IF_q(Y;\what{\bm{\beta}}_q)\IF_q^\top(Y;\what{\bm{\beta}}_q)\}.
\]
Additionally, it is known that $\bm{B}_n^{-1}\bm{A}_n\bm{B}_n^{-1} - \bm{\mathcal{F}}_n^{-1}$, 
is a positive semidefinite matrix. That is, the $\what{\bm{\beta}}_q$ estimator has asymptotic covariance
matrix always larger than the covariance of $\what{\bm{\beta}}_{\ML}$ for the working model
defined in \eqref{eq:density} and \eqref{eq:linear}. To quantify the loss in efficiency
of the maximum L$q$-likelihood estimator, is to consider the asymptotic relative efficiency
of $\what{\bm{\beta}}_q$ with respect to $\what{\bm{\beta}}_{\ML}$, which is defined as,
\begin{equation}\label{eq:are}
  \ARE(\what{\bm{\beta}}_q, \what{\bm{\beta}}_{\ML}) = \tr(\bm{\mathcal{F}}_n^{-1})/\tr(\bm{B}_n^{-1}
  \bm{A}_n\bm{B}_n^{-1}).
\end{equation}
Hence, $\ARE(\what{\bm{\beta}}_q, \what{\bm{\beta}}_{\ML}) \geq 1$. This leads to a procedure for
the selection of the distortion parameter $q$. Following the same reasoning as \cite{Windham:1995}
an alternative is to choose $q$ from the available data as that $q_{\opt}$ value that minimizes
$\tr(\bm{B}_n^{-1}\bm{A}_n\bm{B}_n^{-1})$. In other words, the interest is to select the model for which
the loss in efficiency is the smallest. This procedure has also been used in \cite{Qin:2017}
for adaptive selection of the distortion parameter.

In this paper we follow the guidelines given by \cite{LaVecchia:2015} who proposed a
data-driven method for the selection of $q$ \citep[see also][]{Ribeiro:2023}. In this
procedure the authors introduced the concept of stability condition as a mechanism for
determining the distortion parameter by considering an equally spaced grid, say $1 \geq
q_1 > q_2 > \cdots > q_m$ and for each value in the grid $\what{\bm{\beta}}_{q_j}$, for $j=1,
\dots,m$, is computed. To achieve robustness we can select an optimal value $q_{\opt}$,
as the greatest value $q_j$ satisfying $QV_j < \rho$, where $QV_j = \|\what{\bm{\beta}}_{q_j}
- \what{\bm{\beta}}_{q_{j+1}}\|$ with $\rho$ a threshold value. Based on experimental results,
\cite{LaVecchia:2015} proposed to construct a grid between $q_1 = 1$, and some minimum
value, $q_m$ with increments of 0.01, whereas $\rho = 0.05\,\|\what{\bm{\beta}}_{q_m}\|$.
One of the main advantages of their procedure is that, in the absence of contamination,
it allows to select the distortion parameter as close to 1 as possible, thas is, this
leads to the maximum likelihood estimator. Yet another alternative that has shown remarkable 
performance for $q$-selection based on parametric bootstrap is reported by \cite{Gimenez:2022a,
Gimenez:2022b}.

\begin{remark}
  An additional constraint that the selected value for $q$ must satisfy, which follows
  from the definition for the function $J_q(\theta)$ is that $q\theta$ must be in the
  parameter space $\Theta$. This allows, for example, to refine the grid of search values
  for the procedures proposed by \cite{LaVecchia:2015} or \cite{Gimenez:2022a}.
\end{remark}

\section{Goodness of fit and hypothesis testing}\label{sec:inference}

\subsection{Robust deviance and model selection}

The goodness of fit in generalized linear models, considering the estimation method
based on the L$q$-likelihood function, requires appropriate modifications to obtain
robust extensions of the deviance function. Consider the parameter vector to be partitioned
as $\bm{\beta} = (\bm{\beta}_1^\top,\bm{\beta}_2^\top)^\top$ and suppose that our interest 
consists of testing the hypothesis $H_0:\bm{\beta}_1 = \bm{0}$, where $\bm{\beta}_1\in\Rset^r$. 
Thus, we can evaluate the discrepancy between the model defined by the null hypothesis versus 
the model under $H_1:\bm{\beta}_1\neq \bm{0}$ using the proposal of \cite{Qin:2017}, as
\begin{align}\label{eq:deviance}
  D_q(\bm{Y},\what{\bm{\mu}}) & = 2\big\{L_q(\what{\bm{\beta}}_q) - L_q(\widetilde{\bm{\beta}}_q)\big\} \nonumber \\
  & = 2\sum_{i=1}^n \big\{l_q(f(y_i;k(\bm{x}_i^\top\what{\bm{\beta}}_q),\phi)) - l_q(f(y_i;k(\bm{x}_i^\top
  \widetilde{\bm{\beta}}_q),\phi))\big\},
\end{align}
where $\what{\bm{\beta}}_q$ and $\widetilde{\bm{\beta}}_q$ are the corrected maximum 
L$q$-likelihood estimates for $\bm{\beta}$ obtained under $H_0$ and $H_1$, respectively, 
and using these estimates we have that $\widetilde{\mu}_i = \mu_i(\widetilde{\bm{\beta}}_q)$ 
and $\what{\mu}_i = \mu_i(\what{\bm{\beta}}_q)$, for $i=1,\dots,n$. \cite{Qin:2017} based 
on results available in \cite{Cantoni:2001}, proved that the $D_q(\bm{Y},\what{\bm{\mu}})$ 
statistic has an asymptotic distribution following a discrete mixture of chi-squared random 
variables with one degree of freedom. We should note that the correction term proposed in 
\cite{Qin:2017} in our case zeroed for the surrogate parameter. 

Following \cite{Ronchetti:1997}, we can introduce the Akaike information criterion
based on the L$q$-estimation procedure, as follows
\[
  \AIC_q = -2\sum_{i=1}^n l_q(f(y_i;k(\bm{x}_i^\top\what{\bm{\beta}}_q),\phi))
  + 2\tr(\what{\bm{B}}_n^{-1}\what{\bm{A}}_n),
\]
where $\what{\bm{\beta}}_q$ corresponds to the maximum L$q$-likelihood estimator obtained
from the algorithm defined in \eqref{eq:IWLS}. Using the definitions of $\bm{A}_n$ and 
$\bm{B}_n$, given in \eqref{eq:variability} and \eqref{eq:sensitivity} leads to,
\[
  \AIC_q = -2\sum_{i=1}^n l_q(f(y_i;k(\bm{x}_i^\top\what{\bm{\beta}}_q),\phi)) + \frac{2}{2-q}
  \tr\{(\bm{X}^\top \what{\bm{W}}\what{\bm{J}}\what{\bm{G}}\what{\bm{K}}\bm{X})^{-1}
  \bm{X}^\top\what{\bm{W}}\what{\bm{J}}\bm{X}\}.
\]
Evidently, when the canonical $\theta$-link is considered, we obtain that the penalty
term assumes the form $2\tr(\what{\bm{B}}_n^{-1}\what{\bm{A}}_n) = 2p/(2-q)$ and in the 
case that $q\to 1$ we recover the usual Akaike information criterion, $\AIC$ for model 
selection in the context of maximum likelihood estimation. Some authors \citep[see for instance,]
[]{Ghosh:2016}, have suggested using the Akaike information criterion $\AIC_q$ as an
alternative mechanism for the selection of the tuning parameter, i.e., the distortion
parameter $q$.

\subsection{Hypothesis testing and residual analysis}

A robust alternative for conducting hypothesis testing in the context of maximum
L$q$-likelihood estimation has been proposed by \cite{Qin:2017}, who studied the
properties of L$q$-likelihood ratio type statistics for simple hypotheses. As suggested
in Equation \eqref{eq:deviance} this type of development allows, for example, the
evaluation of the fitted model. In this section we focus on testing linear hypotheses
of the type
\begin{equation}\label{eq:H0}
  H_0:\bm{H\beta}_0 = \bm{h}, \qquad \textrm{against} \qquad H_1:\bm{H\beta}_0\neq \bm{h},
\end{equation}
where $\bm{H}$ es a known matrix of order $d\times p$, with $\rk(\bm{H}) = d$ $(d\leq p)$ 
and $\bm{h}\in\Rset^d$. Wald, score-type and bilinear form \citep{Crudu:2020} statistics for 
testing the hypothesis in \eqref{eq:H0} are given by the following result.

\begin{proposition}
  Given the assumptions of Properties 24.10 and 24.16 in \cite{Gourieroux:1995}, considering 
  that $\bm{A}_n$ and $\bm{B}_n$ converge almost surely to matrices $\bm{A}$ and $\bm{B}$,
  respectively, and under $H_0$, then the Wald, score-type and bilinear form statistics
  given by
  \begin{align}
    W_n & = n(\bm{H}\what{\bm{\beta}}_q - \bm{h})^\top(\bm{H}\what{\bm{B}}{}^{-1}\what{\bm{A}}
    \what{\bm{B}}{}^{-1}\bm{H}^\top)^{-1}(\bm{H}\what{\bm{\beta}}_q - \bm{h}), \label{eq:Wald} \\
    R_n & = \frac{1}{n}\bm{\Psi}_n^\top(\widetilde{\bm{\beta}}_q)\widetilde{\bm{B}}^{-1}\bm{H}^\top
    (\bm{H}\widetilde{\bm{B}}{}^{-1}\widetilde{\bm{A}}\widetilde{\bm{B}}{}^{-1}\bm{H}^\top)^{-1}
    \bm{H}\widetilde{\bm{B}}^{-1}\bm{\Psi}_n(\widetilde{\bm{\beta}}_q), \quad \textrm{and}\label{eq:Rao} \\
    BF_n & = \bm{\Psi}_n^\top(\widetilde{\bm{\beta}}_q)\widetilde{\bm{B}}^{-1}\bm{H}^\top(\bm{H}
    \what{\bm{B}}{}^{-1}\what{\bm{A}}\what{\bm{B}}{}^{-1}\bm{H}^\top)^{-1}(\bm{H}\what{\bm{\beta}}_q - \bm{h}),
    \label{eq:BF}
  \end{align}
  are asymptotically equivalent and follow a chi-square distribution with $d$ degrees of 
  freedom, where $\widetilde{\bm{\beta}}_q$ and $\what{\bm{\beta}}_q$ represent the corrected 
  maximum L$q$-likelihood estimates for $\bm{\beta}$ under the null and alternative hypothesis, 
  respectively. For the estimates of matrices $\bm{A}$ and $\bm{B}$, the notation is analogous. 
  Therefore, we reject the null hypothesis at a level $\alpha$, if either $W_n$, $R_n$ 
  or $BF_n$ exceeds a $100(1-\alpha)\%$ quantile value of the distribution $\chi^2(d)$.
\end{proposition}

Robustness of the statistics defined in \eqref{eq:Wald}-\eqref{eq:BF} can be characterized
using the tools available in \cite{Heritier:1994}. In addition, it is relatively simple to
extend these test statistics to manipulate nonlinear hypotheses following the results developed
by \cite{Crudu:2020}.

Suppose that our aim is to evaluate the inclusion of a new variable in the regression model.
Following \cite{Wang:1985}, we can consider the added-variable model, which is defined through
the linear predictor $\eta_i = \bm{x}_i^\top\bm{\beta} + z_i\gamma = \bm{f}_i^\top\bm{\delta}$, 
where $\bm{f}_i^\top = (\bm{x}_i^\top,z_i)$ and $\bm{\delta} = (\bm{\beta}^\top,\gamma)^\top$, 
with $\theta_i = k(\eta_i)$, for $i=1,\dots,n$. Based on the above results we present the 
score-type statistic for testing the hypothesis $H_0:\gamma = 0$. Let $\bm{F} = (\bm{X},\bm{z})$ 
be the model matrix for the added-variable model, with $\bm{X} = (\bm{x}_1,\dots,\bm{x}_n)^\top$
and $\bm{z} = (z_1,\dots,z_n)^\top$. That notation, enables us to write the estimation function
associated with the maximum L$q$-likelihood estimation problem as:
\[
  \bm{\Psi}_n(\bm{\delta}) = \begin{pmatrix}
    \bm{\Psi}_\beta(\bm{\delta}) \\
    \bm{\Psi}_\gamma(\bm{\delta})
  \end{pmatrix} = \phi\begin{pmatrix}
    \bm{X}^\top \bm{W}^{1/2}\bm{UV}^{-1/2}(\bm{Y} - \bm{\mu}) \\
    \bm{z}^\top \bm{W}^{1/2}\bm{UV}^{-1/2}(\bm{Y} - \bm{\mu})
  \end{pmatrix},
\]
whereas,
\begin{align*}
  \bm{A}_n(\bm{\delta}) & = \frac{\phi}{2-q}\begin{pmatrix}
    \bm{X}^\top \bm{WJX} & \bm{X}^\top \bm{WJz} \\
    \bm{z}^\top \bm{WJX} & \bm{z}^\top \bm{WJz}
  \end{pmatrix}, \\ 
  \bm{B}_n(\bm{\delta}) & = \phi\begin{pmatrix}
    \bm{X}^\top \bm{WJGKX} & \bm{X}^\top \bm{WJGKz} \\
    \bm{z}^\top \bm{WJGKX} & \bm{z}^\top \bm{WJGKz}
  \end{pmatrix}.
\end{align*}
It is easy to notice that the estimate for $\beta$ under the null hypothesis, $H_0:
\gamma = 0$ can be obtained by the Newton-scoring algorithm given in \eqref{eq:IWLS}.
Details of the derivation of the score-type statistic for testing $H_0:\gamma = 0$ 
are deferred to Appendix \ref{app:subvectors}, where it is obtained
\[
  R_n = \bm{\Psi}_\gamma^\top(\what{\bm{\beta}}_q)\{\cov(\bm{\Psi}_\gamma(\what{\bm{\beta}}_q))\}^{-1}
  \bm{\Psi}_\gamma(\what{\bm{\beta}}_q),
\]
which for the context of the added variable model, takes the form:
\[
  R_n = \frac{(2-q)\{\bm{z}^\top\what{\bm{W}}{}^{1/2}\what{\bm{U}}\what{\bm{V}}{}^{-1/2}
  (\bm{Y} - \what{\bm{\mu}})\}^2}{\phi^{-1}\bm{z}^\top(\bm{I}_n - \what{\bm{W}}\what{\bm{J}}
  \what{\bm{G}}\what{\bm{K}}\what{\bm{P}})\what{\bm{W}}\what{\bm{J}}(\bm{I}_n - \what{\bm{P}}
  \what{\bm{W}}\what{\bm{J}}\what{\bm{G}}\what{\bm{K}})\bm{z}},
\]
where $\bm{P} = \bm{X}(\bm{X}^\top \bm{WJGKX})^{-1}\bm{X}^\top$. Thus, we reject the 
null hypothesis by comparing the value obtained for $R_n$ with some percentile value 
$(1- \alpha)$ of the chi-squared distribution with one degree of freedom. This type of 
development has been used, for example, in \cite{Wei:1994} to propose tools for outlier 
detection considering the mean-shift outlier model in both generalized linear and nonlinear 
regression models for the maximum likelihood framework. It should be noted that the 
mean-shift outlier model can be obtained by considering $\bm{z} = (0,\dots,1,\dots,0)^\top$ 
as the vector of zeros except for an 1 at the $i$th position, which leads to the standardized 
residual.
\[
  t_i = \frac{\sqrt{2-q}\,\what{U}_i(Y_i - \what{\mu}_i)}{\phi^{-1/2}\what{J}_i^{1/2}
  \what{V}_i^{1/2}\sqrt{(1 - \what{g}_i\what{k}_i\what{m}_{ii}) - \what{g}_i\what{k}_i
  (\what{m}_{ii} - \what{g}_i\what{k}_i\what{m}_{ii}^*)}}, \qquad i=1,\dots,n,
\]
where $\what{U}_i = U_i(\what{\bm{\beta}})$, $\what{J}_i = J_q^{-\phi}(\theta_i(\what{\bm{\beta}}))$,
$\what{g}_i = \dot{g}(\what{\theta}_i^*)$, $\what{k}_i = \dot{k}(\what{\eta}_i)$, and
$\what{m}_{ii}$, $\what{m}_{ii}^*$ are, respectively, the diagonal elements of matrices
$\what{\bm{M}} = \what{\bm{P}}\what{\bm{W}}\what{\bm{J}}$ and $\what{\bm{M}}{}^2$, for $i=1,\dots,n$. 
It is straightforward to note that $t_i$ has an approximately standard normal 
distribution and can be used for the identification of outliers and possibly influential 
observations. Evidently, we have that for $q = 1$ we recover the standardized residual 
for generalized linear models. We have developed the standardized residual, $t_i$ based 
on the type-score statistic defined based on \eqref{eq:Rao}, further alternatives for 
standardized residuals can be constructed by using the Wald and bilinear-form statistics 
defined in Equations \eqref{eq:Wald} and \eqref{eq:BF}, respectively. Additionally, we 
can consider other types of residuals. Indeed, based on Equation \eqref{eq:deviance} we 
can define the deviance residual as,
\[
  r_i^\dev = \sign(y_i - \what{\mu}_i)\sqrt{d(y_i,\what{\mu}_i)}, \qquad i=1,\dots,n,
\]
with $d(y_i,\what{\mu}_i) = 2\{l_q(y_i,\what{\mu}_i) - l_q(y_i,y_i)\}$, the deviance
associated with the $i$th observation. Moreover, it is also possible to consider the
quantile residual \citep{Dunn:1996}, which is defined as
\[
  r_i^\qtl = \Phi^{-1}\{F(y_i;\what{\mu}_i,\phi)\}, \qquad i=1,\dots,n,
\]
where $F(\cdot;\mu,\phi)$ is the cumulative distribution function associated with the
random variable $y$, while $\Phi(\cdot)$ is the cumulative distribution function of the
standard normal.

Throughout this work we have assumed $\phi$ as a fixed niusance parameter. Following
\cite{Gimenez:2022a, Gimenez:2022b}, we can obtain the maximum L$q$-likelihood estimator
$\what{\phi}_q$ by solving the problem,
\[
  \underset{\phi}{\max}\, H_n(\phi), \qquad H_n(\phi) = \sum_{i=1}^n l_q(f(y_i;
  k(\bm{x}_i^\top\what{\bm{\beta}}_q),\phi)),
\]
which corresponds to a profiled L$q$-likelihood function. It should be noted that to
obtain $\what{\phi}_q$ it is sufficient to consider an one-dimensional optimization
procedure.

\section{Numerical experiments}\label{sec:exp}

\subsection{Example}\label{sec:ex}

We revisit the dataset introduced by \cite{Finney:1947} which aims to assess the
effect of the volume and rate of inspired air on transient vasoconstriction in the
skin of the digits. The response is binary, indicating the occurrence or non-occurrence
of vasoconstriction. Following \cite{Pregibon:1981}, a logistic regression model using
the linear predictor $\eta = \beta_1 + \beta_2\log(\text{volume}) + \beta_3\log(\text{rate})$,
was considered. Observations 4 and 18 have been identified as highly influential on
the estimated coefficients obtained by the maximum likelihood method \citep{Pregibon:1981}.
In fact, several authors have highlighted the extreme characteristic of this dataset 
since the removal of these observations brings the model close to indeterminacy \cite[see, 
for instance][Sec.\,4]{Kunsch:1989}. Table \ref{tab:estimation}, reports the parameter 
estimates and their standard errors obtained from three robust procedures. Here CR denotes 
the fit using the method proposed by \cite{Cantoni:2001}, BY indicates the estimation method 
for logistic regression developed by \cite{Bianco:1996} both available in the robustbase 
package \citep{Todorov:2009} from the R software \citep{R:2022}, while ML$q$, denotes the 
procedure proposed in this work where we have selected $q$ using the mechanism described 
in Section \ref{sec:IF}, in which $q = 0.79$ was obtained. Furthermore, the results of the 
maximum likelihood fit are also reported and estimation was carried out using the weighted 
Bianco and Yohai estimator \citep{Croux:2003}. However, since the explanatory variables do 
not contain any outliers the estimates obtained using BY and weighted BY methods are identical 
and therefore they are not reported here.

\begin{table}[!htp]
  \caption{Estimates by maximum likelihood and robust methods for the skin vaso-constriction data.}\label{tab:estimation}
  {\centering
  \begin{tabular}{l r@{.}l r@{.}l r@{.}l r@{.}l r@{.}l r@{.}l} \hline\hline
                      & \multicolumn{4}{c}{Intercept} & \multicolumn{4}{c}{$\log(\text{volume})$} & \multicolumn{4}{c}{$\log(\text{rate})$} \\ \hline
    ML                &  -2&875 &  (1&321) &  5&179 &  (1&865) &  4&562 &  (1&838) \\
    ML*               & -24&581 & (14&020) & 39&550 & (23&244) & 31&935 & (17&758) \\
    ML$q$, $q = 0.79$ &  -5&185 &  (2&563) &  8&234 &  (3&920) &  7&287 &  (3&455) \\
    CR                &  -2&753 &  (1&327) &  4&974 &  (1&862) &  4&388 &  (1&845) \\
    BY                &  -6&852 & (10&040) & 10&734 & (15&295) &  9&364 & (12&770) \\ \hline\hline
    \multicolumn{13}{l}{{\small * With cases 4 and 18 removed.}} \\ 
  \end{tabular}
  }
\end{table}

\begin{figure}[!ht]
  \vskip -1em
  \centering
  \subfigure[]{
    \includegraphics[width = 0.45\linewidth]{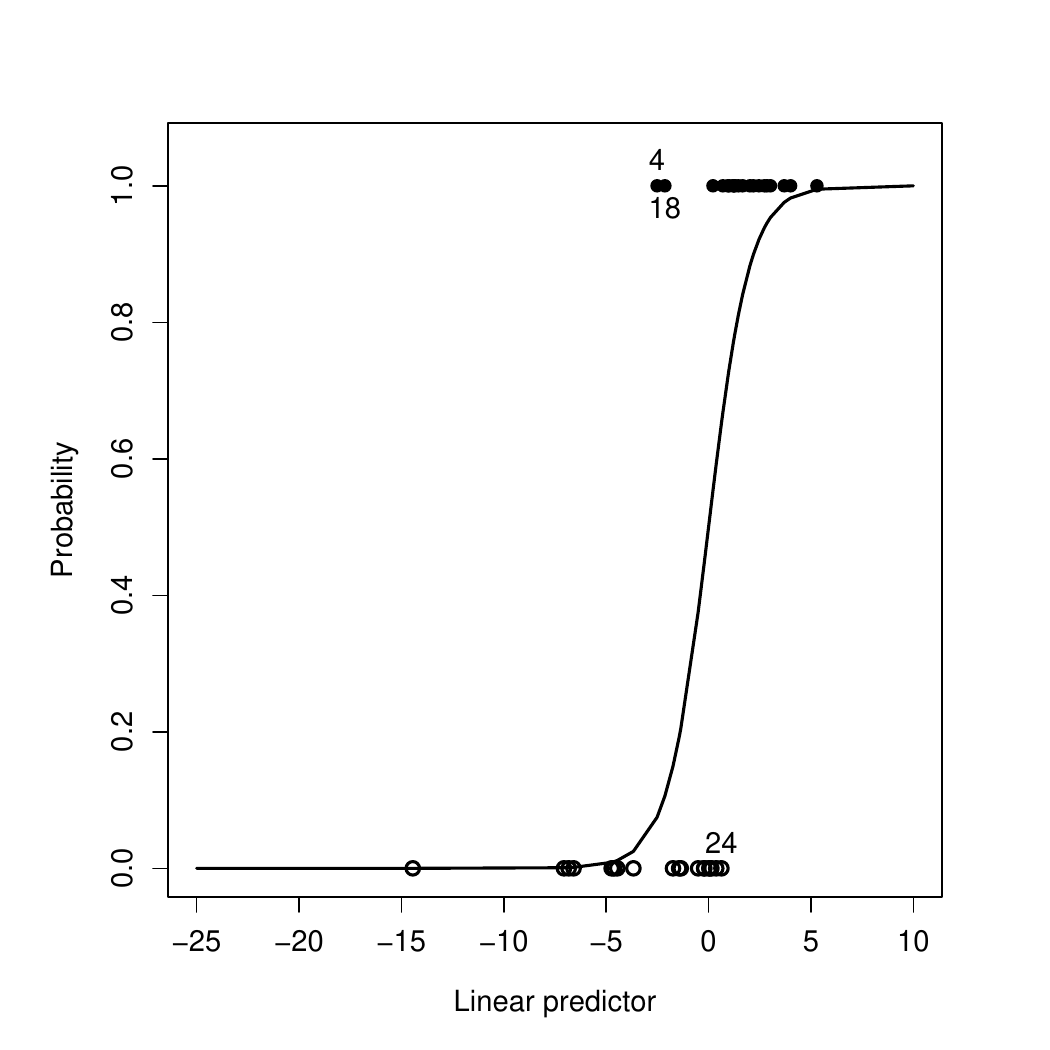}
  }
  \subfigure[]{
    \includegraphics[width = 0.45\linewidth]{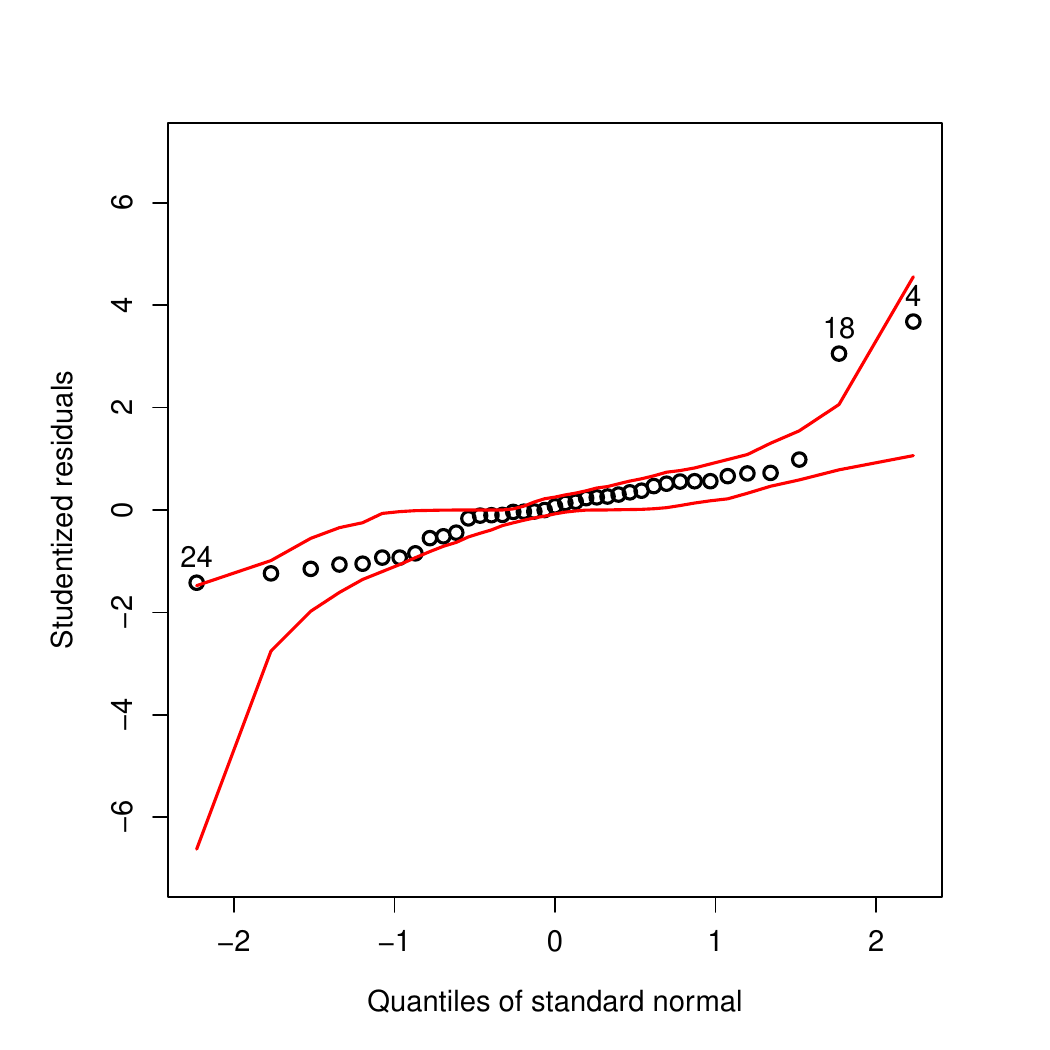}
  }
  \caption{Skin vaso-constriction data: (a) observations and the estimated probabilities 
  and (b) QQ-plot of Studentized residuals with simulation envelope, from the logistic 
  model fitted using maximum likelihood.}\label{fig:ML}
\end{figure}

From Table \ref{tab:estimation} we must highlight the drastic change in the estimates of the 
regression coefficients by maximum likelihood when observations 4 and 18 are deleted \citep[see 
also][]{Pregibon:1981, Wang:1985}. It is interesting to note that the results for this case 
are very close to those offered by the estimation procedure based on the $L_1$-norm introduced 
by \cite{Morgenthaler:1992}. Moreover, as discussed in Sec.\,6.17 of \cite{Atkinson:2000} the 
removal of observations 4, 18 and 24 leads to a perfect fit. Indeed, the elimination of observations 
4 and 18 leads to a huge increment in the standard errors.

\begin{figure}[!ht]
  \vskip -1em
  \centering
  \includegraphics[width = 0.5\linewidth]{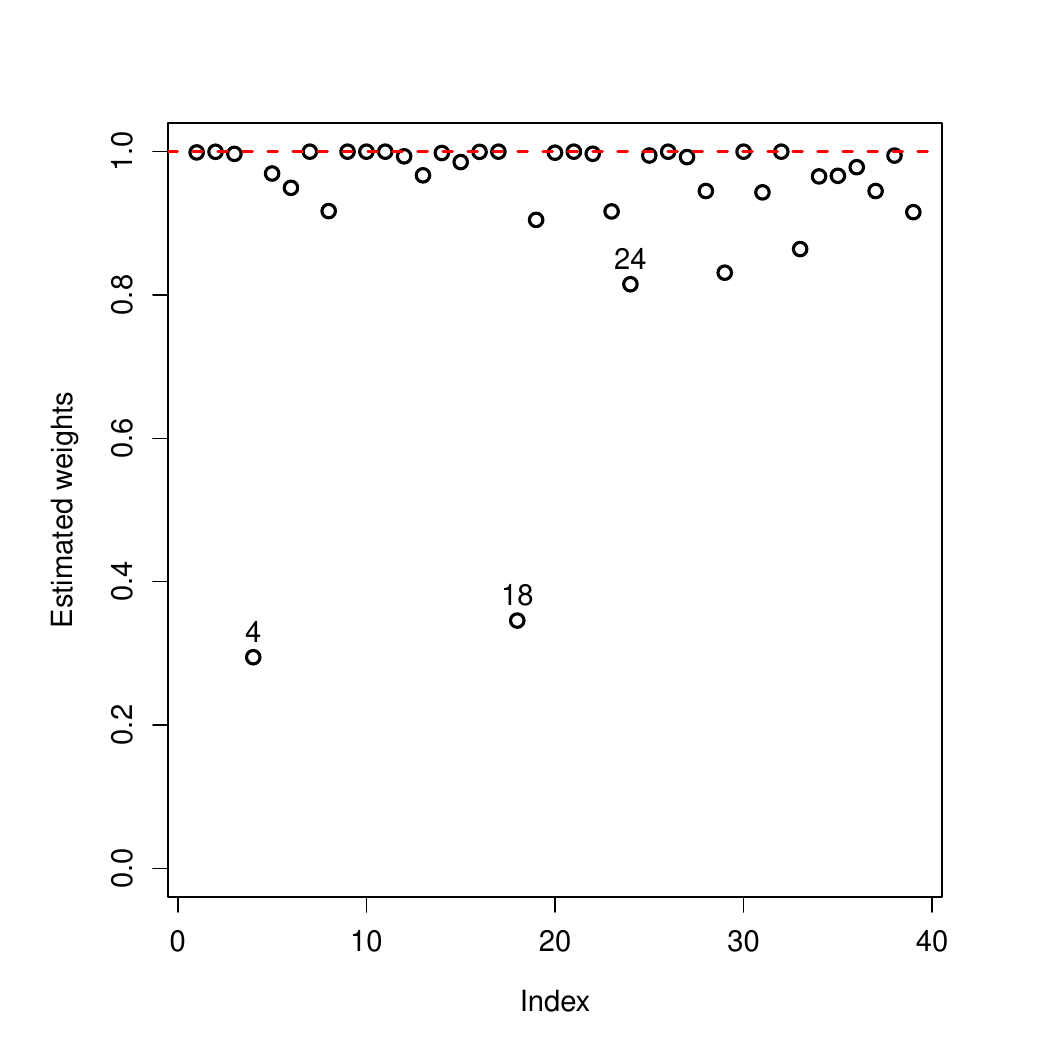}
  \caption{Skin vaso-constriction data: weights obtained from the logistic model fitted 
  using maximum L$q$-likelihood with $q = 0.79$.}\label{fig:robweights}
\end{figure}

\begin{figure}[!ht]
  \vskip -1em
  \centering
  \subfigure[]{
    \includegraphics[width = 0.45\linewidth]{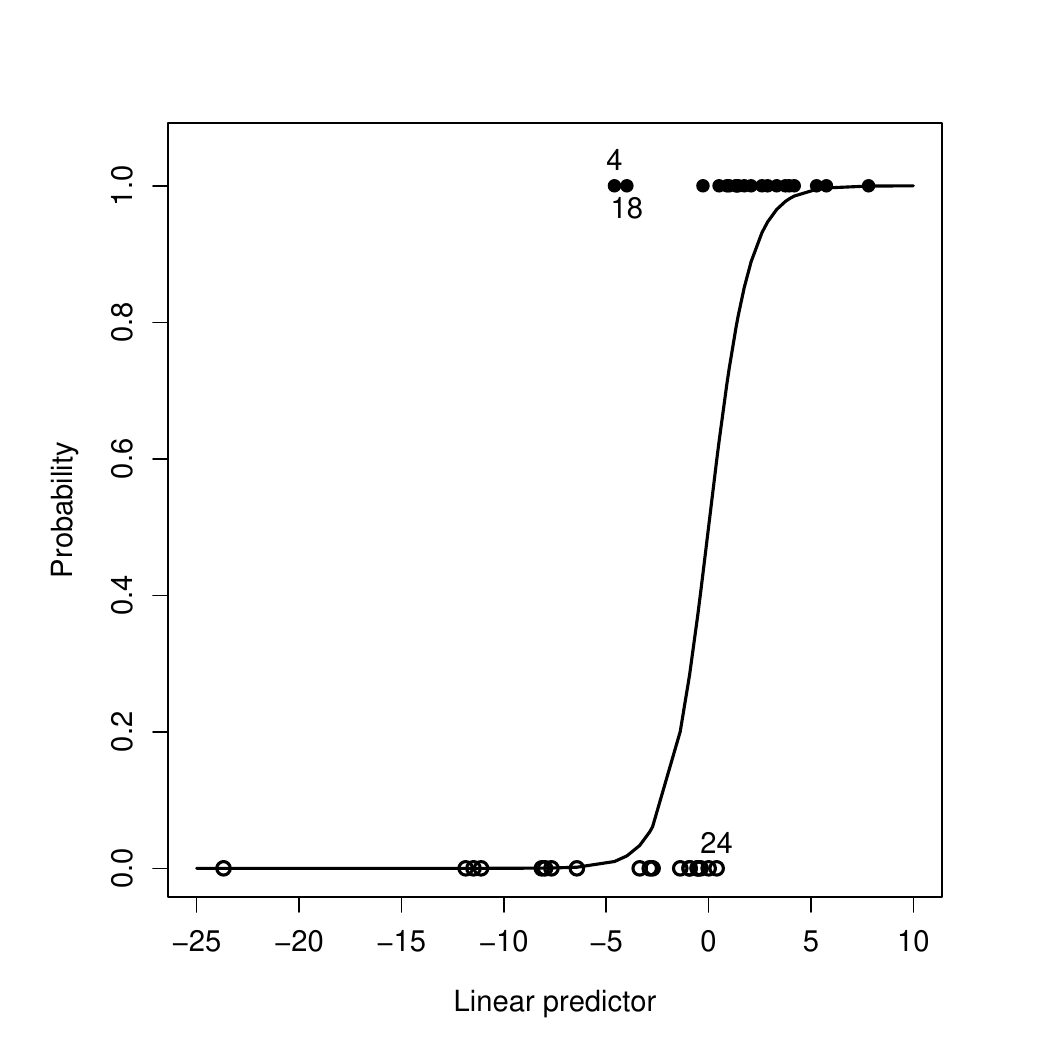}
  }
  \subfigure[]{
    \includegraphics[width = 0.45\linewidth]{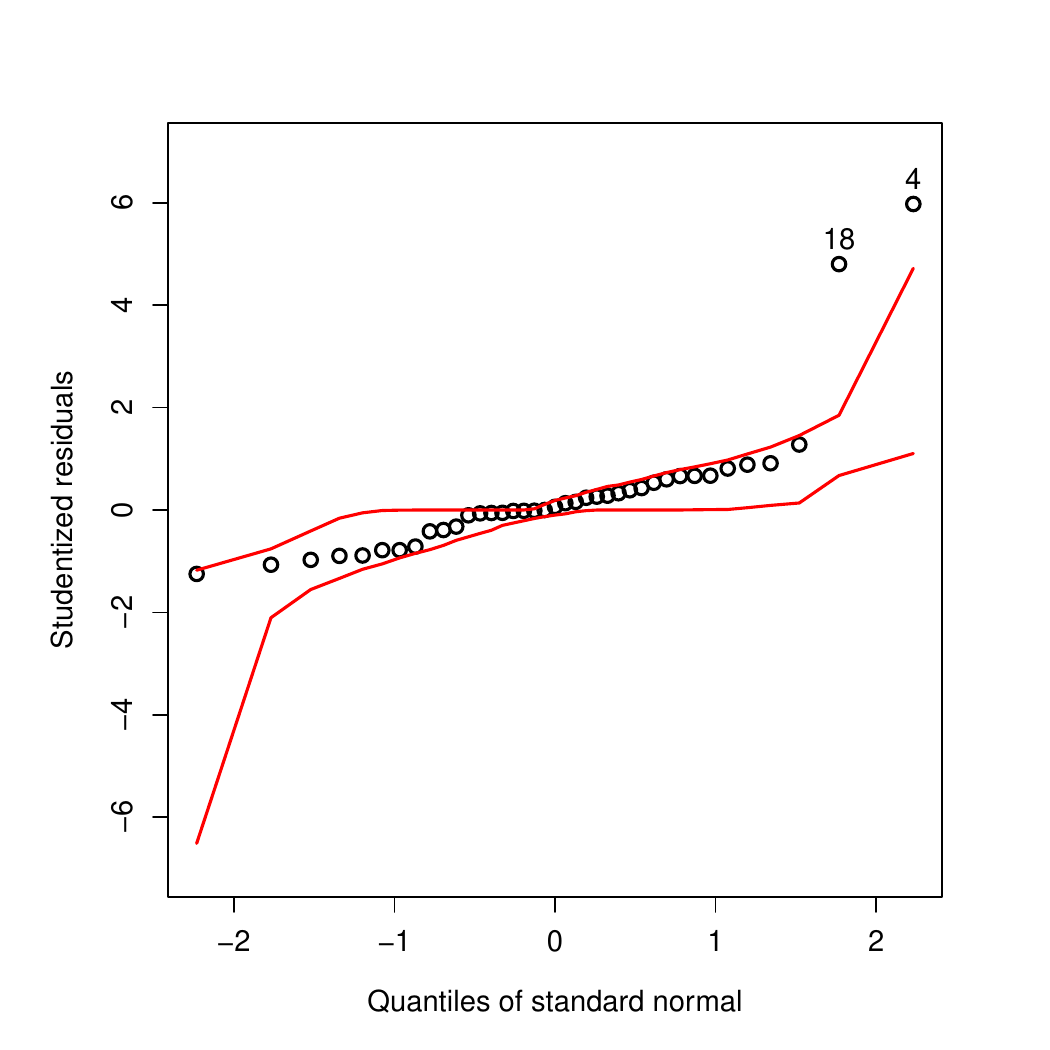}
  }
  \caption{Skin vaso-constriction data: (a) observations and the estimated probabilities 
  and (b) QQ-plot of Studentized residuals with simulation envelope, from the logistic 
  model fitted using maximum L$q$-likelihood, $q = 0.79$.}\label{fig:MLq}
\end{figure}

Figure \ref{fig:ML}\,(a) shows a strong overlap in the groups with zero and unit response. The 
poor fit is also evident from the QQ-plot of Studentized residuals with simulated envelopes (see, 
Figure \ref{fig:ML}\,(b)). The estimated weights $\what{U}_i = U_i(\what{\bm{\beta}})$, $(i=1,\dots,39)$ 
are displayed in Figure \ref{fig:robweights}, where it is possible to appreciate the ability of 
the algorithm to downweight observations 4, 18 and 24. It should be stressed that the estimation 
of the regression coefficients by the maximum L$q$-likelihood method with $q = 0.79$ is very similar 
to that obtained by the resistant procedure discussed by \cite{Pregibon:1982}. Figure \ref{fig:MLq}, 
reveals that the maximum L$q$-likelihood procedure offers a better fit and the QQ-plot with simulated 
envelope correctly identifies observations 4 and 18 as outliers.

In Appendix \ref{app:additional} the fit for the proposed model using values ranging from 
$q = 1$ to 0.74, is reported. Table \ref{tab:grid} shows that as the value of $q$ decreases, 
there is an increase in the standard error of the estimates. Additionally, Figures \ref{fig:wts} 
and \ref{fig:prob} from Appendix \ref{app:additional} show how as $q$ decreases the weights 
of observations 4 and 18 decrease rapidly to zero. Indeed, when $q \leq 0.76$ the problem of 
indeterminacy underlying these data becomes evident.

\subsection{Simulation study}

To evaluate the performance of the proposed estimation procedure we developed a small 
Monte Carlo simulation study. We consider a Poisson regression model with logarithmic 
link function, based on the linear predictor
\begin{equation}\label{eq:simul}
  \eta_i = \beta_1 x_{i1} + \beta_2 x_{i2} + \beta_3 x_{i3}, \qquad i=1,\dots,n,
\end{equation}
where $x_{ij}\sim \mathsf{U}(0,1)$ for $j=1,2,3$ and $\bm{\beta} = (\beta_1,\beta_2,\beta_3)^\top = 
(1,1,1)^\top$. We construct $M = 1000$ data sets with sample size $n = 25, 100$ and 
$400$ from the model \eqref{eq:simul}. Following \cite{Cantoni:2006} the observations 
are contaminated by multiplying by a factor of $\nu$ a percentage $\varepsilon$ of randomly 
selected responses. We have considered $\nu = 2,5$ and $\varepsilon = 0.00, 0.05, 0.10$ 
and $0.25$ as contamination percentages. For contaminated as well as non-contaminated 
data we have carried out the estimation by maximum likelihood, using the method of \cite{Cantoni:2001} 
and the procedure proposed in this paper.

Let $\what{\bm{\beta}}_k = (\what{\beta}_{k1},\dots,\what{\beta}_{kp})^\top$ be the vector of 
estimated parameters for the $k$th simulated sample $(k = 1,\dots,M)$. We summarize the results 
of the simulations by calculating the following statistics \citep{Hosseinian:2011}
\[
  \bias = \Big\|\frac{1}{M}\sum_{k=1}^M (\what{\bm{\beta}}_k - \bm{\beta})\Big\|, \qquad 
  \IQR = \frac{1}{p}\sum_{j=1}^p \IQR(\what{\beta}_{kj}),
\]
where $\IQR$ denotes the interquartile range and for our study $p = 3$.
\begin{table}[!htp]
  \caption{Bias of the ML$q$ estimators for various values of the distortion parameter $q$ under different levels of contamination.}\label{tab:bias}
  {\centering
  \begin{tabular}{rcccccccccccc} \hline\hline
    $n$ & $\varepsilon$ & $\nu$ & \multicolumn{9}{c}{$q$} & CR \\ \cline{4-12}
        &      &   & 1.00  & 0.99  & 0.97  & 0.95  & 0.93  & 0.91  & 0.89  & 0.87  & 0.85  & \\ \hline
     25 & 0.05 & 2 & 0.020 & 0.007 & 0.039 & 0.076 & 0.113 & 0.151 & 0.188 & 0.225 & 0.262 & 0.009 \\
    100 &      &   & 0.042 & 0.023 & 0.020 & 0.058 & 0.097 & 0.135 & 0.173 & 0.211 & 0.248 & 0.024 \\
    400 &      &   & 0.043 & 0.023 & 0.018 & 0.057 & 0.096 & 0.135 & 0.173 & 0.211 & 0.248 & 0.023 \\
     25 & 0.05 & 5 & 0.114 & 0.085 & 0.050 & 0.064 & 0.098 & 0.141 & 0.182 & 0.217 & 0.244 & 0.023 \\
    100 &      &   & 0.155 & 0.101 & 0.024 & 0.053 & 0.100 & 0.137 & 0.154 & 0.160 & 0.158 & 0.031 \\
    400 &      &   & 0.161 & 0.101 & 0.015 & 0.049 & 0.099 & 0.142 & 0.182 & 0.202 & 0.195 & 0.029 \\
     25 & 0.10 & 2 & 0.055 & 0.037 & 0.021 & 0.051 & 0.089 & 0.128 & 0.166 & 0.204 & 0.242 & 0.030 \\
    100 &      &   & 0.083 & 0.061 & 0.021 & 0.028 & 0.067 & 0.107 & 0.147 & 0.187 & 0.226 & 0.048 \\
    400 &      &   & 0.084 & 0.063 & 0.019 & 0.023 & 0.065 & 0.106 & 0.146 & 0.186 & 0.225 & 0.048 \\
     25 & 0.10 & 5 & 0.221 & 0.178 & 0.097 & 0.009 & 0.082 & 0.141 & 0.172 & 0.184 & 0.184 & 0.058 \\
    100 &      &   & 0.285 & 0.213 & 0.093 & 0.037 & 0.044 & 0.089 & 0.150 & 0.177 & 0.171 & 0.065 \\
    400 &      &   & 0.296 & 0.216 & 0.088 & 0.013 & 0.052 & 0.141 & 0.301 & 0.301 & 0.245 & 0.063 \\
     25 & 0.25 & 2 & 0.187 & 0.166 & 0.123 & 0.083 & 0.051 & 0.046 & 0.072 & 0.106 & 0.143 & 0.131 \\
    100 &      &   & 0.194 & 0.171 & 0.124 & 0.078 & 0.034 & 0.022 & 0.061 & 0.098 & 0.123 & 0.130 \\
    400 &      &   & 0.199 & 0.175 & 0.127 & 0.079 & 0.033 & 0.014 & 0.056 & 0.101 & 0.126 & 0.131 \\
     25 & 0.25 & 5 & 0.590 & 0.530 & 0.410 & 0.424 & 0.459 & 0.468 & 0.484 & 0.503 & 0.504 & 0.229 \\
    100 &      &   & 0.591 & 0.518 & 0.407 & 0.741 & 0.750 & 0.702 & 0.649 & 0.597 & 0.544 & 0.190 \\
    400 &      &   & 0.610 & 0.530 & 0.401 & 0.838 & 0.783 & 0.730 & 0.676 & 0.623 & 0.568 & 0.187 \\ \hline\hline
  \end{tabular}
  }
\end{table}

\begin{table}
  \caption{Interquartile range of the ML$q$ estimators for various values of the distortion parameter $q$ under different levels of contamination.}\label{tab:iqr}
  {\centering
  \begin{tabular}{rcccccccccccc} \hline\hline
    $n$ & $\varepsilon$ & $\nu$ & \multicolumn{9}{c}{$q$} & CR \\ \cline{4-12}
        &      &   & 1.00  & 0.99  & 0.97  & 0.95  & 0.93  & 0.91  & 0.89  & 0.87  & 0.85  & \\ \hline
     25 & 0.05 & 2 & 0.431 & 0.426 & 0.413 & 0.401 & 0.393 & 0.386 & 0.377 & 0.369 & 0.363 & 0.425 \\
    100 &      &   & 0.195 & 0.190 & 0.183 & 0.178 & 0.175 & 0.171 & 0.167 & 0.164 & 0.161 & 0.187 \\
    400 &      &   & 0.103 & 0.100 & 0.095 & 0.091 & 0.090 & 0.089 & 0.087 & 0.085 & 0.083 & 0.097 \\
     25 & 0.05 & 5 & 0.611 & 0.572 & 0.492 & 0.442 & 0.432 & 0.426 & 0.440 & 0.437 & 0.454 & 0.444 \\
    100 &      &   & 0.363 & 0.303 & 0.227 & 0.201 & 0.185 & 0.178 & 0.182 & 0.190 & 0.220 & 0.192 \\
    400 &      &   & 0.195 & 0.160 & 0.118 & 0.103 & 0.097 & 0.093 & 0.090 & 0.095 & 0.121 & 0.101 \\
     25 & 0.10 & 2 & 0.454 & 0.449 & 0.435 & 0.423 & 0.413 & 0.404 & 0.393 & 0.388 & 0.386 & 0.453 \\
    100 &      &   & 0.201 & 0.196 & 0.191 & 0.187 & 0.182 & 0.175 & 0.171 & 0.165 & 0.163 & 0.190 \\
    400 &      &   & 0.109 & 0.106 & 0.103 & 0.100 & 0.097 & 0.094 & 0.093 & 0.092 & 0.090 & 0.101 \\
     25 & 0.10 & 5 & 0.807 & 0.753 & 0.619 & 0.548 & 0.547 & 0.578 & 0.619 & 0.689 & 0.730 & 0.474 \\
    100 &      &   & 0.421 & 0.358 & 0.282 & 0.236 & 0.234 & 0.296 & 0.418 & 0.459 & 0.475 & 0.204 \\
    400 &      &   & 0.221 & 0.193 & 0.150 & 0.126 & 0.117 & 0.263 & 0.229 & 0.220 & 0.241 & 0.111 \\
     25 & 0.25 & 2 & 0.519 & 0.513 & 0.506 & 0.496 & 0.483 & 0.472 & 0.473 & 0.475 & 0.472 & 0.520 \\
    100 &      &   & 0.239 & 0.238 & 0.230 & 0.223 & 0.215 & 0.210 & 0.205 & 0.204 & 0.208 & 0.224 \\
    400 &      &   & 0.118 & 0.116 & 0.114 & 0.112 & 0.110 & 0.107 & 0.106 & 0.104 & 0.107 & 0.114 \\
     25 & 0.25 & 5 & 1.008 & 1.003 & 1.057 & 1.293 & 1.303 & 1.298 & 1.221 & 1.167 & 1.139 & 0.712 \\
    100 &      &   & 0.465 & 0.462 & 0.469 & 0.560 & 0.502 & 0.492 & 0.481 & 0.470 & 0.459 & 0.288 \\
    400 &      &   & 0.226 & 0.228 & 0.244 & 0.240 & 0.241 & 0.236 & 0.231 & 0.226 & 0.220 & 0.142 \\ \hline\hline
  \end{tabular}
  }
\end{table}

Tables \ref{tab:bias} and \ref{tab:iqr} present the simulation results. It should be noted 
that, in the presence of contamination, the maximum L$q$-likelihood estimation method with 
values of the distortion parameter $q$ around 0.99 to 0.95 outperforms the procedure proposed 
by \cite{Cantoni:2001} with the exception of severe levels of contamination (i.e. $\varepsilon 
= 0.25$ and $\nu = 5$). These results are in concordance with Theorem 3.1 described in \cite{Ferrari:2010}, 
although they highlight the need to understand in detail the proportion of aberrant data supported 
by this estimation procedure. The simulations also reveal the trade-off for the selection of 
the distortion parameter, while decreasing the value of $q$ tends to choose models whose standard 
error turns out to be smaller, introduces considerable bias and therefore we recommend the selection 
procedures proposed by \cite{LaVecchia:2015} and \cite{Ribeiro:2023}. Indeed, for non-contaminated 
data (results not presented here) such procedure correctly leads to select the maximum likelihood 
estimation method, i.e. $q = 1$.

\section{Concluding remarks}\label{sec:conc}

The methodology described in this work provides a fully parametric robust estimation mechanism that depends 
on a single tuning parameter $q$ which controls the robustness and efficiency of the procedure. The simplicity 
of the proposed approach as well as its interesting asymptotic properties have also allowed us to introduce 
measures to evaluate the goodness of fit of the model, to carry out hypothesis tests as well as to define 
standardized residuals. In particular, the estimation can be characterized by iteratively weighted least 
squares which allows to re-use existing code, resulting in a procedure less sensitive to outlying observations. 
We have described several strategies for the selection of the distortion (tuning) parameter. In particular, 
the proposal of \cite{LaVecchia:2015} represents a very low computational cost alternative that seems to work 
well in the application with real data. The Monte Carlo simulation study reveals a satisfactory performance 
of the procedure based on maximum L$q$-likelihood estimation in presence of contamination even when compared 
to some popular GLM alternatives for robust estimation. It should be stressed that despite the robustness of 
the proposed procedure, there may still be observations that exert disproportionate influence on key aspects 
of the statistical model; the study to assess the local influence of this type of observations is a topic for 
further research which is being developed by the authors. An implementation of the ML$q$ estimation for GLM 
along with the experiments developed in this work are publicly available on github.

\section*{Code and software availability}

All analyses and simulations were conducted in the R environment for statistical computing. The replication files 
related to this article are available online at \url{https://github.com/faosorios/robGLM}

\section*{Acknowledgements}

The authors were partially supported by the UTFSM grant PI\_LI\_19\_11. Additionally, authors 
acknowledge the support of the computing infrastructure of the Applied Laboratory of Spatial 
Analysis UTFSM - ALSA (MT7018).


\appendix
\section{Fisher consistency}\label{app:Fisher}

We say that a random variable $Y$ belongs to the exponential family of distributions
if its probability density function is given by
\begin{equation}\label{eq:A-density}\tag{A.1}
  f(y;\theta,\phi) = \exp[\phi\{y\theta - b(\theta)\} + c(y,\phi)],
\end{equation}
where $b(\cdot)$ and $c(\cdot,\cdot)$ are some specific functions, $\theta$ is the natural
parameter defined in $\Theta\subset\Rset$ and $\phi > 0$ is a dispersion parameter. If a
random variable has a density function \eqref{eq:A-density}, we shall denote $Y\sim\mathsf{FE}(\theta,\phi)$.
To verify Fisher consistency, the following Lemma is introduced,

\begin{lemma}\label{lem:expectation}
  Suppose $Y\sim\mathsf{FE}(\theta,\phi)$ with expectation $E(Y) = \dot{b}(\theta)$,
  and consider $U(\theta) = \{f(y;\theta,\phi)\}^{1-q}$ for $q > 0$. Then, for an integrable
  function $h(Y;\theta)$, it follows that
  \[
    \E_0\{U^r(\theta)h(Y;\theta)\} = J_q^{-\phi}(\theta)\E_q\{h(Y;\theta)\},
  \]
  where $\theta_0 = q\theta$, and $J_q(\theta) = \exp\{-qb(\theta) + b(q\theta)\}$ is defined
  in \cite{Menendez:2000}, whereas the expectation $\E_q\{h(Y;\theta)\}$ must be calculated based
  on the density function
  \begin{equation}\label{eq:q-density}\tag{A.2}
    f_q(y;\theta,\phi) = \exp[\phi(r(1 - q) + q)\{y\theta - b(\theta)\} + c_{q,r}(y,\phi)],
  \end{equation}
  and $c_{q,r}(y,\phi) = (r(1 - q) + 1)c(y,\phi)$.
\end{lemma}

\begin{proof}
  It is desired to compute expectations of the type
  \[
    \E_0\{U^r(\theta)h(Y;\theta)\} = \int_{\mathcal{Y}} U^r(\theta)h(y;\theta)
    f(y;\theta_0,\phi){{\rm d}y}.
  \]
  For $\theta_0 = q\theta$, we have
  \begin{align*}
    U^r(\theta) f(y;\theta_0,\phi) = {} & \{f(y;\theta,\phi)\}^{r(1 - q)} f(y;q\theta,\phi) \\
    = {} & \exp[r(1 - q)\phi\{y\theta - b(\theta)\} + r(1 - q)c(y;\phi)] \\
    {} & \times\exp[\phi\{yq\theta - b(q\theta)\} + c(y,\phi)].
  \end{align*}
  It is easy to notice that
  \begin{align*}
    r(1 - q)\phi\{y\theta & - b(\theta)\} + \phi\{yq\theta - b(q\theta)\} \\
    & = \phi(r(1 - q) + q)\{y\theta - b(\theta)\} + \phi\{qb(\theta) - b(q\theta)\}.
  \end{align*}
  Thus,
  \[
    U^r(\theta)f(y;\theta_0,\phi) = \exp[\phi\{qb(\theta) - b(q\theta)\}]
    \exp[\phi(r(1 - q) + q)\{y\theta - b(\theta)\} + c_q(y;\phi)],
  \]
  where $c_q(y,\phi) = (r(1 - q) + 1)c(y,\phi)$ and let
  \[
    J_q(\theta) = \exp\{-qb(\theta) + b(q\theta)\}, \qquad q\theta \in \Theta,
  \]
  \citep[see][]{Menendez:2000}, which leads to write
  \[
    \E_0\{U^r(\theta)h(Y;\theta)\} = J_q^{-\phi}(\theta)\E_q\{h(Y,\theta)\},
  \]
  because $E_q\{h(Y,\theta)\} = \int_{\mathcal{Y}}h(y;\theta) f_q(y;\theta,\phi){{\rm d}y}$,
  with $f_q(y;\theta,\phi)$ as defined in \eqref{eq:q-density}.
\end{proof}

\smallskip

\begin{proposition}\label{prop:consistency}
  The estimation function defined in Equation \eqref{eq:psi-GLM} is unbiased for
  the surrogate parameter $\theta_* = \theta_0/q$.
\end{proposition}

\begin{proof}
  We have to calculate the expectation,
  \[
    \E_0\{\bm{\Psi}_n(\bm{\beta}_*)\} = \phi\sum_{i=1}^n W_i^{1/2}\frac{\E_0\{U_i(\bm{\beta}_*)
    (Y_i - \mu_{*,i})\}}{\sqrt{V_i}}\bm{x}_i.
  \]
  Since $U_i(\bm{\beta}_*) = \{f(y_i;\theta_i(\bm{\beta}_*),\phi)\}^{1-q}$, it is enough to
  obtain $\E_0\{U(\theta_*)(Y - \mu_*)\}$. Using Lemma \ref{lem:expectation},
  leads to
  \[
    \E_0\{U(\theta_*)(Y - \mu_*)\} = J_q^{-\phi}(\theta_*)\E_q(Y - \mu_*).
  \]
  Noticing that
  \[
    \E_q(Y - \mu_*) = \int_{\mathcal{Y}} \{y - \dot{b}(\theta_*)\}f_q(y;\theta_*,\phi){{\rm d}y}
    = 0,
  \]
  yields $\E_0\{\bm{\Psi}_n(\bm{\beta}_*)\} = 0$. That is, it is an unbiased estimation function
  for $\theta_*$.
\end{proof}

\section{Corrected version of the maximum L$q$-likelihood estimator in GLM}

Let $\theta_0$ be the true parameter, \cite{Ferrari:2010} highlighted that
for $q$ fixed, the maximum L$q$-likelihood estimator $\hat{\theta}_q^*$ converges in probability
to $\theta_* = \theta_0/q$. Here, $\theta_*$ is called the surrogate parameter of $\theta_0$.
This leads to the correction $\hat{\theta}_q = q\hat{\theta}_q^*\stackrel{p}{\to}
q\theta_* = \theta_0$. Considering the $\theta$-link, that is $\theta = k(\eta)$, we write
$\bm{\eta}_*=k^{-1}(\bm{\theta}_*)=\bm{X\beta}_*$ for the corresponding surrogates. Then we have that
\[
  \hat{\eta}_q = k^{-1}(q\hat{\theta}_q^*) \stackrel{p}{\to} k^{-1}(\theta_0) = \eta_0.
\]
This leads to the following correction for the linear predictor $\hat{\eta}_q = k^{-1}
(qk(\hat{\eta}_q^*))$, which yields the calibration function in Equation \eqref{eq:calibration}.

\section{Asymptotic covariance}

Previous to the calculation of matrices defined in Equations \eqref{eq:variability} and \eqref{eq:sensitivity}, 
we introduce the following Lemma.

\begin{lemma}\label{lem:variance}
  Consider $Y\sim\mathsf{FE}(\theta,\phi)$ and assume the elements given in Lemma \ref{lem:expectation}.
  Therefore it is straightforward that
  \[
    \E_0\{U^r(\theta)(Y - \mu)^2\} = \frac{\phi^{-1}}{r(1 - q) + q}
    J_q^{-\phi}(\theta) V(\mu),
  \]
  where $V(\mu) = \ddot{b}(\theta)$ is the variance function.
\end{lemma}

\begin{proof}
  Using Lemma \ref{lem:expectation}, we have that $\E_0\{U^r(\theta)(Y - \mu)^2\} = J_q^{-\phi}(\theta)
  \E_q\{(Y - \mu)^2\}$. Since,
  \[
    \E_q\{(Y - \mu)^2\} = \frac{\phi^{-1}}{r(1 - q) + q}\,\ddot{b}(\theta),
  \]
  the result follows.
\end{proof}

\smallskip

\begin{proof}[Proof of Equation \eqref{eq:variability}]
  We have that
  \[
    \bm{A}_n(\bm{\beta}) = \E_0\{\bm{\Psi}_n(\bm{\beta})\bm{\Psi}_n^\top(\bm{\beta})\} 
    = \sum_{i=1}^n \E_0\{U_i^2(\bm{\beta})\bm{s}_i(\bm{\beta})\bm{s}_i^\top(\bm{\beta})\},
  \]
  where $\bm{s}_i(\bm{\beta}) = \phi\{y_i - \dot{b}(\theta_i)\}\dot{k}(\eta_i)\bm{x}_i$. Thus,
  \begin{align}\label{eq:var}\tag{C.1}
    \bm{A}_n(\bm{\beta}) = \phi^2\sum_{i=1}^n \E_0[U_i^2(\bm{\beta})\{y_i - \dot{b}(\theta_i)\}^2]
    \{\dot{k}(\eta_i)\}^2\bm{x}_i\bm{x}_i^\top.
  \end{align}
  Using Lemma \ref{lem:variance} with $r = 2$, we obtain
  \[
    \E_0[U_i^2(\bm{\beta})\{y_i - \dot{b}(\theta_i)\}^2] = \frac{\phi^{-1}}{2(1 - q) + q}\,
    J_q^{-\phi}(\theta_i)\ddot{b}(\theta_i).
  \]

  Substituting this result in Equation \eqref{eq:var}, it follows that
  \begin{align*}
    \bm{A}_n(\bm{\beta}) & = \frac{\phi}{2 - q}\sum_{i=1}^n J_q^{-\phi}(\theta_i)\ddot{b}(\theta_i)
    \{\dot{k}(\eta_i)\}^2 \bm{x}_i\bm{x}_i^\top = \frac{\phi}{2 - q}\sum_{i=1}^n W_i J_q^{-\phi}(\theta_i)
    \bm{x}_i\bm{x}_i^\top \\
    & = \frac{\phi}{2-q}\,\bm{X}^\top \bm{WJX},
  \end{align*}
  this yields Equation \eqref{eq:variability}.
\end{proof}

\smallskip

\begin{proof}[Proof of Equation \eqref{eq:sensitivity}]
  To obtain $\bm{B}_n(\bm{\beta})$, first consider
  \[
    \frac{\partial}{\partial\eta_i} U_i(\eta_i)s_i(\eta_i) = \frac{\partial U_i(\eta_i)}
    {\partial\eta_i} s_i(\eta_i) + U_i(\eta_i)\frac{\partial s_i(\eta_i)}{\partial\eta_i},
  \]
  with
  \begin{align*}
    \frac{\partial U_i(\eta_i)}{\partial\eta_i} & = (1 - q)\{f(y_i;\eta_i,\phi)\}^{1-q-1}
    \frac{\partial f(y_i;\eta_i,\phi)}{\partial\eta_i} \\
    & = (1 - q)U_i(\eta_i)s_i(\eta_i),
  \end{align*}
  where $s_i(\eta_i) = \phi\{y_i - \dot{b}(\theta_i)\}\dot{k}(\eta_i)$. We also have that
  \[
    \frac{\partial s_i(\eta_i)}{\partial\eta_i} = -\phi\,\ddot{b}(\theta_i)
    \{\dot{k}(\eta_i)\}^2 + \phi\{y_i - \dot{b}(\theta_i)\}\ddot{k}(\eta_i).
  \]
  On the one hand,
  \begin{align*}
    \E_0\Big\{-\frac{\partial U_i(\eta_i)}{\partial\eta_i}s_i(\eta_i)\Big\}
    & = -\E_0\{(1-q)U_i(\eta_i)s_i^2(\eta_i)\} \\
    & = -(1-q)\phi^2 \E_0[U_i(\eta_i)\{y_i - \dot{b}(\theta_i)\}^2]\{\dot{k}(\eta_i)\}^2.
  \end{align*}
  We known that $\E_0[U_i(\eta_i)\{y_i - \dot{b}(\theta_i)\}^2] = \phi^{-1}J_q^{-\phi}(\theta_i)
  \ddot{b}(\theta_i)$, therefore
  \begin{equation}\label{eq:s1}\tag{C.2}
    \E_0\Big\{-\frac{\partial U_i(\eta_i)}{\partial\eta_i}s_i(\eta_i)\Big\}
    = -(1-q)\phi J_q^{-\phi}(\theta_i)\ddot{b}(\theta_i)\{\dot{k}(\eta_i)\}^2,
  \end{equation}
  on the other hand,
  \[
    \E_0\Big\{-U_i(\eta_i)\frac{\partial s_i(\eta_i)}{\partial\eta_i}\Big\}
    = \phi\ddot{b}(\theta_i)\{\dot{k}(\eta_i)\}^2 \E_0\{U_i(\eta_i)\}
    - \phi \E_0[U_i(\eta_i)\{y_i - \dot{b}(\theta_i)\}]\ddot{k}(\eta_i).
  \]
  By using Lemma S1, we have that
  \[
    \E_0\{U_i(\eta_i)\} = J_q^{-\phi}(\theta_i), \qquad
    \E_0[U_i(\eta_i)\{y_i - \dot{b}(\theta_i)\}] = 0.
  \]
  Thus,
  \begin{equation}\label{eq:s2}\tag{C.3}
    \E_0\Big\{-U_i(\eta_i)\frac{\partial s_i(\eta_i)}{\partial\eta_i}\Big\}
    = \phi\ddot{b}(\theta_i)\{\dot{k}(\eta_i)\}^2 J_q^{-\phi}(\theta_i).
  \end{equation}
  Using Equations \eqref{eq:s1} and \eqref{eq:s2}, yields to the expectation
  \[
    \E_0\Big\{-\frac{\partial}{\partial\eta_i} U_i(\eta_i)s_i(\eta_i)\Big\}
    = q\phi J_q^{-\phi}(\theta_i)\ddot{b}(\theta_i)\{\dot{k}(\eta_i)\}^2
    = q\phi J_q^{-\phi}(\theta_i) W_i
  \]
  We known that $\partial\eta_i^*/\partial\eta_i = q^{-1}\partial k^{-1}(\theta_i^*)/\partial
  \theta_i^*\,\dot{k}(\eta_i)$. Therefore,
  \[
    \E_0\Big\{-\frac{\partial\bm{\Psi}_n(\bm{\beta})}{\partial\bm{\beta}^\top}\Big\}
    = \phi\sum_{i=1}^n J_q^{-\phi}(\theta_i) W_i \frac{\partial k^{-1}(\theta_i^*)}
    {\partial\theta_i^*}\dot{k}(\eta_i)\bm{x}_i\bm{x}_i^\top.
  \]
  Considering $\bm{G} = \diag(\dot{g}(\theta_1^*),\dots,\dot{g}(\theta_n^*))$ with
  $g(\theta_i^*) = k^{-1}(\theta_i^*)$ and $\bm{K} = \diag(\dot{k}(\theta_1),\dots,$
  $\dot{k}(\theta_n))$, we obtain
  \[
    \E_0\Big\{-\frac{\partial\bm{\Psi}_n(\bm{\beta})}{\partial\bm{\beta}^\top}\Big\}
    = \phi \bm{X}^\top \bm{WJGKX},
  \]
  and the proof is complete.
\end{proof}

\section{L$q$-likelihood based tests}

\begin{proof}[Proof of Proposition 2]
  Following Property 24.16 in \cite{Gourieroux:1995}, we know that
  \[
    \sqrt{n}(\what{\bm{\beta}}_q - \bm{\beta}_0) \stackrel{D}{\longrightarrow} 
    \mathsf{N}_p(\bm{0},\bm{B}^{-1}\bm{AB}^{-1}).
  \]
  This leads to,
  \[
    \sqrt{n} \bm{H}(\what{\bm{\beta}}_q - \bm{\beta}_0) \stackrel{D}{\longrightarrow} 
    \mathsf{N}_k(\bm{0},\bm{HB}^{-1}\bm{AB}^{-1}\bm{H}^\top).
  \]
  Evidently, $\sqrt{n}\bm{H}(\what{\bm{\beta}}_q - \bm{\beta}_0) = \sqrt{n}(\bm{H}
  \what{\bm{\beta}}_q - \bm{H\beta}_0)$. Now, under $H_0$, we have $\bm{H\beta}_0 = \bm{h}$. 
  Hence
  \begin{equation}\label{eq:asymp}\tag{D.1}
    \sqrt{n} (\bm{H}\what{\bm{\beta}}_q - \bm{h}) \stackrel{D}{\longrightarrow} 
    \mathsf{N}_k(\bm{0},\bm{HB}^{-1}\bm{AB}^{-1}\bm{H}^\top).
  \end{equation}
  Thus, under $H_0:\bm{H\beta}_0 = \bm{h}$, we obtain
  \[
    W_n = n(\bm{H}\what{\bm{\beta}}_q - \bm{h})^\top(\bm{H}\what{\bm{B}}^{-1}\what{\bm{A}}
    \what{\bm{B}}^{-1}\bm{H}^\top)^{-1}(\bm{H}\what{\bm{\beta}}_q - \bm{h}) \stackrel{D}{\longrightarrow} 
    \chi^2(k),
  \]
  which allows us to establish Equation \eqref{eq:Wald}.

  \smallskip

  To establish \eqref{eq:Rao}, note that the constrained estimator $\widetilde{\bm{\beta}}_q$ is defined
  as solution of the Lagrangian problem,
  \[
    Q_n(\bm{\beta},\bm{\lambda}) = L_q(\bm{\beta}) + \bm{\lambda}^\top(\bm{h} - \bm{H\beta}),
  \]
  where $\bm{\lambda}$ denotes a $k\times 1$ vector of Lagrange multipliers. We have that
  $\widetilde{\bm{\beta}}_q$ and $\widetilde{\bm{\lambda}}_n$ satisfy the first order conditions 
  \citep[see Property 24.10 of][]{Gourieroux:1995},
  \begin{equation}\label{eq:FO}\tag{D.2}
    \bm{\Psi}_n(\widetilde{\bm{\beta}}_q) - \bm{H}^\top\widetilde{\bm{\lambda}}_n = 0, \qquad
    \bm{H}\widetilde{\bm{\beta}}_q - \bm{h} = \bm{0},
  \end{equation}
  and the corrected $\widetilde{\bm{\beta}}_q$ is consistent. Considering Taylor expansions of
  $\bm{\Psi}_n(\what{\bm{\beta}}_q)$ and $\bm{\Psi}_n(\widetilde{\bm{\beta}}_q)$ around $\bm{\beta}$, 
  assuming that $\bm{B}_n \stackrel{a.s.}{\longrightarrow} \bm{B}$ uniformly and following simple 
  calculations yield
  \[
    \sqrt{n}(\what{\bm{\beta}}_q - \widetilde{\bm{\beta}}_q) = \bm{B}^{-1}\,\frac{1}{\sqrt{n}}
    \bm{\Psi}_n(\widetilde{\bm{\beta}}_q) + o_p(\bm{1}).
  \]
  From the first order condition \eqref{eq:FO}, leads to
  \[
    \sqrt{n} \bm{H}(\what{\bm{\beta}}_q - \widetilde{\bm{\beta}}_q) = \bm{HB}^{-1}\bm{H}^\top\,
    \frac{\widetilde{\bm{\lambda}}_n}{\sqrt{n}} + o_p(\bm{1}).
  \]
  Moreover, using $\bm{H}\widetilde{\bm{\beta}}_q - \bm{h} = \bm{0}$, we find
  \begin{equation}\label{eq:diff}\tag{D.3}
    \sqrt{n} (\bm{H}\what{\bm{\beta}}_q - \bm{h}) - \sqrt{n} (\bm{H}\widetilde{\bm{\beta}}_q - \bm{h}) 
    = \sqrt{n} \bm{H}(\what{\bm{\beta}}_q - \widetilde{\bm{\beta}}_q)
  \end{equation}
  From \eqref{eq:asymp} and \eqref{eq:diff}, we obtain
  \begin{equation}\label{eq:hyp}\tag{D.4}
    \sqrt{n} \bm{H}(\what{\bm{\beta}}_q - \widetilde{\bm{\beta}}_q) = \sqrt{n} (\bm{H}\what{\bm{\beta}}_q 
    - \bm{h}) \stackrel{D}{\longrightarrow} \mathsf{N}_k(\bm{0},\bm{HB}^{-1}\bm{AB}^{-1}\bm{H}^\top).
  \end{equation}
  Furthermore, Equation \eqref{eq:diff} enables us to write,
  \[
    \frac{\widetilde{\bm{\lambda}}_n}{\sqrt{n}} = (\bm{HB}^{-1}\bm{H}^\top)^{-1}\sqrt{n}(\bm{H}\what{\bm{\beta}}_q 
    - \bm{h}) + o_p(\bm{1}).
  \]
  Then, using \eqref{eq:hyp}, it follows that
  \begin{equation}\label{eq:lambda}\tag{D.5}
    \frac{\widetilde{\bm{\lambda}}_n}{\sqrt{n}} \stackrel{D}{\longrightarrow} \mathsf{N}_k(\bm{0},(\bm{HB}^{-1}
    \bm{H}^\top)^{-1}\bm{HB}^{-1}\bm{AB}^{-1}\bm{H}^\top(\bm{HB}^{-1}\bm{H}^\top)^{-1}).
  \end{equation}
  This result allows us to define the statistic
  \begin{align*}
    R_n & = \frac{1}{n}\,\widetilde{\bm{\lambda}}_n^\top \bm{H}\widetilde{\bm{B}}^{-1}\bm{H}^\top
    (\bm{H}\widetilde{\bm{B}}^{-1}\widetilde{\bm{A}}\widetilde{\bm{B}}^{-1}\bm{H}^\top)^{-1}\bm{H}
    \widetilde{\bm{B}}^{-1}\bm{H}^\top\widetilde{\bm{\lambda}}_n \\
    & = \frac{1}{n}\,\bm{\Psi}_n^\top(\widetilde{\bm{\beta}}_q)\widetilde{\bm{B}}^{-1}\bm{H}^\top
    (\bm{H}\widetilde{\bm{B}}^{-1}\widetilde{\bm{A}}\widetilde{\bm{B}}^{-1}\bm{H}^\top)^{-1}\bm{H}
    \widetilde{\bm{B}}^{-1}\bm{\Psi}_n(\widetilde{\bm{\beta}}_q) \\
    & \stackrel{D}{\longrightarrow} \chi^2(k),
  \end{align*}
  as desired.

  \smallskip

  Let $\bm{HB}^{-1}\bm{AB}^{-1}\bm{H}^\top = \bm{RR}^\top$ where $\bm{R}$ is a nonsingular $k\times k$ matrix. 
  Thus, it follows from \eqref{eq:asymp} and \eqref{eq:lambda},
  \begin{align*}
    \sqrt{n}\,\bm{R}^{-1}(\bm{H}\what{\bm{\beta}}_q - \bm{h}) & \stackrel{D}{\longrightarrow} \mathsf{N}_k(\bm{0},\bm{I}_k) \\
    \bm{R}^{-1}\bm{HB}^{-1}\bm{H}^\top\,\frac{\widetilde{\bm{\lambda}}_n}{\sqrt{n}} & \stackrel{D}{\longrightarrow} 
    \mathsf{N}_k(\bm{0},\bm{I}_k),
  \end{align*}
  and this imply that,
  \begin{align*}
    BF_n & = \Big\{\bm{R}^{-1}\bm{H}\widetilde{\bm{B}}^{-1}\bm{H}^\top\,\frac{\widetilde{\bm{\lambda}}_n}{\sqrt{n}}\Big\}^\top
    \sqrt{n}\,\bm{R}^{-1}(\bm{H}\what{\bm{\beta}}_q - \bm{h}) \\
    & = \widetilde{\bm{\lambda}}_n^\top \bm{H}\widetilde{\bm{B}}^{-1}\bm{H}^\top \bm{R}^{-\top}\bm{R}^{-1}
    (\bm{H}\what{\bm{\beta}}_q - \bm{h}) \\[.35em]
    & = \widetilde{\bm{\lambda}}_n^\top \bm{H}\widetilde{\bm{B}}^{-1}\bm{H}^\top(\bm{H}\what{\bm{B}}^{-1}\what{\bm{A}}
    \what{\bm{B}}^{-1}\bm{H}^\top)^{-1}(\bm{H}\what{\bm{\beta}}_q - \bm{h}) \stackrel{D}{\longrightarrow} \chi^2(k),
  \end{align*}
  which can be written alternatively, as
  \[
    BF_n = \bm{\Psi}_n^\top(\widetilde{\bm{\beta}}_q)\widetilde{\bm{B}}^{-1}\bm{H}^\top(\bm{H}\what{\bm{B}}^{-1}
    \what{\bm{A}}\what{\bm{B}}^{-1}\bm{H}^\top)^{-1}(\bm{H}\what{\bm{\beta}}_q - \bm{h}) \stackrel{D}{\longrightarrow} 
    \chi^2(k),
  \]
  and the proposition is verified.
\end{proof}

\section{Score-type statistic for hypotheses about subvectors}\label{app:subvectors}

Consider that the hypothesis $H_0:\bm{\beta}_2 = \bm{\beta}_{02}$ with $\bm{\beta} = (\bm{\beta}_1^\top,
\bm{\beta}_2^\top)^\top$ where $\bm{\beta}_1\in\Rset^r$ and $\bm{\beta}_2\in\Rset^{p-r}$. Define the Lagrangian 
function,
\[
  Q_n(\bm{\beta}) = L_q(\bm{\beta}) + \bm{\lambda}^\top(\bm{\beta}_{02} - \bm{\beta}_2),
\]
where $\bm{\lambda}\in\Rset^{p-r}$ is a vector of Lagrange multipliers. Let $\bm{\Psi}_1(\bm{\beta})
= \partial L_q(\bm{\beta})/\partial\bm{\beta}_1$ and $\bm{\Psi}_2(\bm{\beta}) = \partial L_q(\bm{\beta})/\partial\bm{\beta}_2$ 
the derivatives of $L_q(\bm{\beta})$ with respect to $\bm{\beta}_1$ and $\bm{\beta}_2$, respectively. 
The first order condition leads to,
\[
  \bm{\Psi}_1(\widetilde{\bm{\beta}}) = \bm{0}, \qquad \bm{\Psi}_2(\widetilde{\bm{\beta}}) 
  - \widetilde{\bm{\lambda}} = \bm{0}, \qquad \bm{\beta}_{02} - \bm{\beta}_2 = \bm{0},
\]
where $\widetilde{\bm{\beta}} = (\widetilde{\bm{\beta}}_1^\top,\widetilde{\bm{\beta}}_2^\top)^\top$ and 
$\widetilde{\bm{\lambda}}$ denote the constrained estimators of $\bm{\beta}$ and $\bm{\lambda}$.

\medskip

Following \cite{Boos:1992}, we can consider a Taylor expansion of $\bm{\Psi}_n(\widetilde{\bm{\beta}})$ around
$\bm{\beta}_0 = (\bm{\beta}_{01}^\top,\bm{\beta}_{02}^\top)^\top$ as
\begin{align*}
  \bm{\Psi}_1(\widetilde{\bm{\beta}}) & = \bm{\Psi}_1(\bm{\beta}_0) + \frac{\partial\bm{\Psi}_1(\bm{\beta}_0)}
  {\partial\bm{\beta}_1^\top}(\widetilde{\bm{\beta}}_1 - \bm{\beta}_{01}) + o_p(\bm{1}), \\
  \bm{\Psi}_2(\widetilde{\bm{\beta}}) & = \bm{\Psi}_2(\bm{\beta}_0) + \frac{\partial\bm{\Psi}_2(\bm{\beta}_0)}
  {\partial\bm{\beta}_1^\top}(\widetilde{\bm{\beta}}_1 - \bm{\beta}_{01}) + o_p(\bm{1}).
\end{align*}
However, $\bm{\Psi}_1(\widetilde{\bm{\beta}}) = \bm{0}$ and substituting $-\partial\bm{\Psi}_1(\bm{\beta}_0)/\partial\bm{\beta}_1^\top$
and $-\partial\bm{\Psi}_2(\bm{\beta}_0)/\partial\bm{\beta}_1^\top$ by their expectations, follow that
\begin{align}
  \bm{\Psi}_1(\widetilde{\bm{\beta}}) & = \bm{\Psi}_1(\bm{\beta}_0) - \bm{B}_{11}(\bm{\beta}_0)
  (\widetilde{\bm{\beta}}_1 - \bm{\beta}_{01}) + o_p(\bm{1}), \label{eq:Psi1}\tag{E.1} \\
  \bm{\Psi}_2(\widetilde{\bm{\beta}}) & = \bm{\Psi}_2(\bm{\beta}_0) - \bm{B}_{21}(\bm{\beta}_0)
  (\widetilde{\bm{\beta}}_1 - \bm{\beta}_{01}) + o_p(\bm{1}). \label{eq:Psi2}\tag{E.2}
\end{align}
with $\bm{B}_{11}(\bm{\beta}) = \E_0\{-\partial\bm{\Psi}_1(\bm{\beta})/\partial\bm{\beta}_1^\top\}$ and 
$\bm{B}_{21}(\bm{\beta}) = \E_0\{-\partial\bm{\Psi}_2(\bm{\beta})/\partial\bm{\beta}_1^\top\}$. Noticing 
that
\begin{equation}\label{eq:diff-b0}\tag{E.3}
  \widetilde{\bm{\beta}}_1 - \bm{\beta}_{01} = \bm{B}_{11}^{-1}(\bm{\beta}_0)\bm{\Psi}_1(\bm{\beta}_0) + o_p(\bm{1}).
\end{equation}
Substituting Equation \eqref{eq:diff-b0} into \eqref{eq:Psi2}, we obtain
\begin{align*}
  \bm{\Psi}_2(\widetilde{\bm{\beta}}) & = \bm{\Psi}_2(\bm{\beta}_0) - \bm{B}_{21}(\bm{\beta}_0)\bm{B}_{11}^{-1}(\bm{\beta}_0)
  \bm{\Psi}_1(\bm{\beta}_0) + o_p(\bm{1}) \\
  & = (-\bm{B}_{21}(\bm{\beta}_0)\bm{B}_{11}^{-1}(\bm{\beta}_0), \bm{I})\begin{pmatrix}
    \bm{\Psi}_1(\bm{\beta}_0) \\
    \bm{\Psi}_2(\bm{\beta}_0)
  \end{pmatrix} + o_p(\bm{1}).
\end{align*}
Moreover,
\[
  \cov(\bm{\Psi}_2(\widetilde{\bm{\beta}})) = (-\bm{B}_{21}(\bm{\beta}_0)\bm{B}_{11}^{-1}(\bm{\beta}_0), \bm{I})
  \cov(\bm{\Psi}_n(\widetilde{\bm{\beta}}))
  \begin{pmatrix}
    -\bm{B}_{11}^{-\top}(\bm{\beta}_0)\bm{B}_{21}^\top(\bm{\beta}_0) \\
    \bm{I}
  \end{pmatrix}.
\]
This leads to the score-type statistic
\[
  R_n = \bm{\Psi}_2^\top(\widetilde{\bm{\beta}})\{\cov(\bm{\Psi}_2(\widetilde{\bm{\beta}}))\}^{-1}
  \bm{\Psi}_2(\widetilde{\bm{\beta}}).
\]

\subsection{Score-type statistic for adding a variable}

Assume that $\eta_i = \bm{x}_i^\top\bm{\beta} + z_i\gamma = \bm{f}_i^\top\bm{\delta}$, where 
$\bm{f}_i = (\bm{x}_i^\top,z_i)$ and $\bm{\delta} = (\bm{\beta}^\top,\gamma)^\top$. Let $\bm{F} 
= (\bm{X},\bm{z})$ be the model matrix for the added-variable
model, where
\[
  \bm{X} = \begin{pmatrix}
    \bm{x}_1^\top \\
    \vdots \\
    \bm{x}_n^\top
  \end{pmatrix}, \qquad \bm{z} = \begin{pmatrix}
    z_1 \\
    \vdots \\
    z_n
  \end{pmatrix}.
\]
Thus,
\[
  \bm{\Psi}_n(\delta) = \begin{pmatrix}
    \bm{\Psi}_\beta(\bm{\delta}) \\
    \bm{\Psi}_\gamma(\bm{\delta})
  \end{pmatrix} = \phi\begin{pmatrix}
    \bm{X}^\top \bm{W}^{1/2}\bm{UV}^{-1/2}(\bm{Y} - \bm{\mu}) \\
    \bm{z}^\top \bm{W}^{1/2}\bm{UV}^{-1/2}(\bm{Y} - \bm{\mu})
  \end{pmatrix}.
\]
We have that,
\begin{align*}
  \bm{A}_n(\delta) & = \frac{\phi}{2-q}\bm{F}^\top \bm{WJF} = \frac{\phi}{2-q}\begin{pmatrix}
    \bm{X}^\top \bm{WJX} & \bm{X}^\top \bm{WJz} \\
    \bm{z}^\top \bm{WJX} & \bm{z}^\top \bm{WJz}
  \end{pmatrix} \\
  \bm{B}_n(\delta) & = \phi \bm{F}^\top \bm{WJGKF} = \phi\begin{pmatrix}
    \bm{X}^\top \bm{WJGKX} & \bm{X}^\top \bm{WJGKz} \\
    \bm{z}^\top \bm{WJGKX} & \bm{z}^\top \bm{WJGKz}
  \end{pmatrix}.
\end{align*}
Consider the following partition for matrices $\bm{A}_n(\bm{\delta})$ and $\bm{B}_n(\bm{\delta})$,
\[
  \bm{A}_n(\delta) = \begin{pmatrix}
    \bm{A}_{11} & \bm{A}_{12} \\
    \bm{A}_{21} & \bm{A}_{22}
  \end{pmatrix}, \qquad \bm{B}_n(\delta) = \begin{pmatrix}
    \bm{B}_{11} & \bm{B}_{12} \\
    \bm{B}_{21} & \bm{B}_{22}
  \end{pmatrix}.
\]
Therefore, the covariance of $\bm{\Psi}_\gamma(\bm{\delta})$ assumes the form,
\[
  \cov(\bm{\Psi}_\gamma(\bm{\delta})) = \bm{A}_{22} - \bm{A}_{21}\bm{B}_{11}^{-1}\bm{B}_{12} 
  - (\bm{B}_{21}\bm{B}_{11}^{-1}\bm{A}_{12} - \bm{B}_{21}\bm{B}_{11}^{-1}\bm{A}_{11}\bm{B}_{11}^{-1}
  \bm{B}_{12}).
\]
After some simple algebra, we obtain
\[
  \cov(\bm{\Psi}_\gamma(\bm{\delta})) = \frac{\phi}{2-q}\bm{z}^\top(\bm{I} - \bm{WJGKP})
  \bm{WJ}(\bm{I} - \bm{PWJGK})\bm{z},
\]
where $\bm{P} = \bm{X}(\bm{X}^\top \bm{WJGKX})^{-1}\bm{X}^\top$. This, leads to
\begin{align*}
  R_n & = \Big(\frac{\phi}{2-q}\Big)^{-1}\frac{\{\phi \bm{z}^\top\what{\bm{W}}^{1/2}
  \what{\bm{U}}\what{\bm{V}}^{-1/2}(\bm{Y} - \bm{\mu})\}^2}{\bm{z}^\top(\bm{I} - \what{\bm{W}}
  \what{\bm{J}}\what{\bm{G}}\what{\bm{K}}\what{\bm{P}})\what{\bm{W}}\what{\bm{J}}(\bm{I} -
  \what{\bm{P}}\what{\bm{W}}\what{\bm{J}}\what{\bm{G}}\what{\bm{K}})\bm{z}} \\
  & = \frac{(2-q)\{\bm{z}^\top\what{\bm{W}}^{1/2}\what{\bm{U}}\what{\bm{V}}^{-1/2}(\bm{Y} 
  - \bm{\mu})\}^2}{\phi^{-1}\bm{z}^\top(\bm{I} - \what{\bm{W}}\what{\bm{J}}\what{\bm{G}}
  \what{\bm{K}}\what{\bm{P}})\what{\bm{W}}\what{\bm{J}}(\bm{I} - \what{\bm{P}}\what{\bm{W}}
  \what{\bm{J}}\what{\bm{G}}\what{\bm{K}})\bm{z}} \stackrel{D}{\longrightarrow} \chi^2(1).
\end{align*}

The foregoing allows us to note that,
\[
  \frac{\sqrt{2-q}\{\bm{z}^\top\what{\bm{W}}^{1/2}\what{\bm{U}}\what{\bm{V}}^{-1/2}
  (\bm{Y} - \bm{\mu})\}}{\phi^{-1/2}\sqrt{\bm{z}^\top(\bm{I} - \what{\bm{W}}\what{\bm{J}}
  \what{\bm{G}}\what{\bm{K}}\what{\bm{P}})\what{\bm{W}}\what{\bm{J}}(\bm{I} - \what{\bm{P}}
  \what{\bm{W}}\what{\bm{J}}\what{\bm{G}}\what{\bm{K}})\bm{z}}} \stackrel{D}{\longrightarrow} 
  \mathsf{N}(0,1).
\]
Now, consider $\bm{z} = (0,\dots,1,\dots,0)^\top$ a vector of zeros except for an 1 at the
$i$th position. It is straightforward to note that,
\[
  \bm{z}^\top\what{\bm{W}}^{1/2}\what{\bm{U}}\what{\bm{V}}^{-1/2}(\bm{Y} - \bm{\mu}) 
  = \what{W}_i^{1/2}\what{U}_i(y_i - \hat{\mu}_i)/\what{V}_i^{1/2}, 
\]
and 
\[
  \bm{z}^\top(\bm{I} - \what{\bm{W}}\what{\bm{J}}\what{\bm{G}}\what{\bm{K}}\what{\bm{P}})
  \what{\bm{W}}\what{\bm{J}}(\bm{I} - \what{\bm{P}}\what{\bm{W}}\what{\bm{J}}\what{\bm{G}}
  \what{\bm{K}})\bm{z} = \what{W}_i\what{J}_i\{(1 - \what{g}_i\what{k}_i\what{m}_{ii}) 
  - \what{g}_i\what{k}_i(\what{m}_{ii} - \what{g}_i\what{k}_i\what{m}_{ii}^*)\}.
\]

This yields to the standardized residual,
\[
  t_i = \frac{\sqrt{2-q}\,\what{W}_i^{1/2}\what{U}_i(y_i - \what{\mu}_i)/\hat{V}_i^{1/2}}
  {\phi^{-1/2}\what{W}_i^{1/2}\what{J}_i^{1/2}\sqrt{(1 - \what{g}_i\what{k}_i\what{m}_{ii})
  - \what{g}_i\what{k}_i(\what{m}_{ii} - \what{g}_i\what{k}_i\what{m}_{ii}^*)}},
\]
which approximately follows an standard normal distribution.

\section{Additional Results}\label{app:additional}

Additional tables and figures related to the example of skin vaso-constriction data
are presented below.

\begin{table}[!ht]
  \caption{Estimates by maximum L$q$-likelihood for the skin vaso-constriction data.}\label{tab:grid}
  \begin{tabular}{r@{.}l r@{.}l r@{.}l r@{.}l r@{.}l r@{.}l r@{.}l} \hline\hline
    \multicolumn{2}{c}{$q$} & \multicolumn{4}{c}{Intercept} & \multicolumn{4}{c}{$\log(\text{volume})$} & \multicolumn{4}{c}{$\log(\text{rate})$} \\ \hline
    1&00  &  -2&875 &  (1&321) &  5&179 &  (1&865) &  4&562 &  (1&838) \\
    0&98  &  -2&919 &  (1&349) &  5&220 &  (1&909) &  4&600 &  (1&875) \\
    0&96  &  -2&970 &  (1&382) &  5&271 &  (1&960) &  4&649 &  (1&918) \\
    0&94  &  -3&033 &  (1&420) &  5&337 &  (2&020) &  4&710 &  (1&967) \\
    0&92  &  -3&109 &  (1&465) &  5&421 &  (2&092) &  4&789 &  (2&026) \\
    0&90  &  -3&205 &  (1&519) &  5&531 &  (2&179) &  4&892 &  (2&097) \\
    0&88  &  -3&327 &  (1&587) &  5&677 &  (2&289) &  5&027 &  (2&185) \\
    0&86  &  -3&488 &  (1&673) &  5&877 &  (2&430) &  5&211 &  (2&297) \\
    0&84  &  -3&712 &  (1&712) &  6&165 &  (2&624) &  5&473 &  (2&450) \\
    0&82  &  -4&047 &  (1&965) &  6&614 &  (2&914) &  5&875 &  (2&677) \\
    0&80  &  -4&636 &  (2&273) &  7&439 &  (3&431) &  6&601 &  (3&078) \\
    0&79* &  -5&185 &  (2&563) &  8&234 &  (3&920) &  7&287 &  (3&455) \\
    0&78  &  -6&322 &  (3&166) &  9&936 &  (4&934) &  8&727 &  (4&233) \\
    0&76  & -18&054 & (11&594) & 28&906 & (19&195) & 23&609 & (14&760) \\
    0&74  & -20&645 & (15&478) & 33&394 & (25&732) & 26&922 & (19&695) \\ \hline\hline
    \multicolumn{14}{l}{{\small * reported in Section \ref{sec:ex}.}} \\ 
  \end{tabular}
\end{table}

\newpage

\begin{figure}[!ht]
  \vskip -1em
  \centering
  \subfigure[$q = 0.98$]{
    \includegraphics[width = 0.28\linewidth]{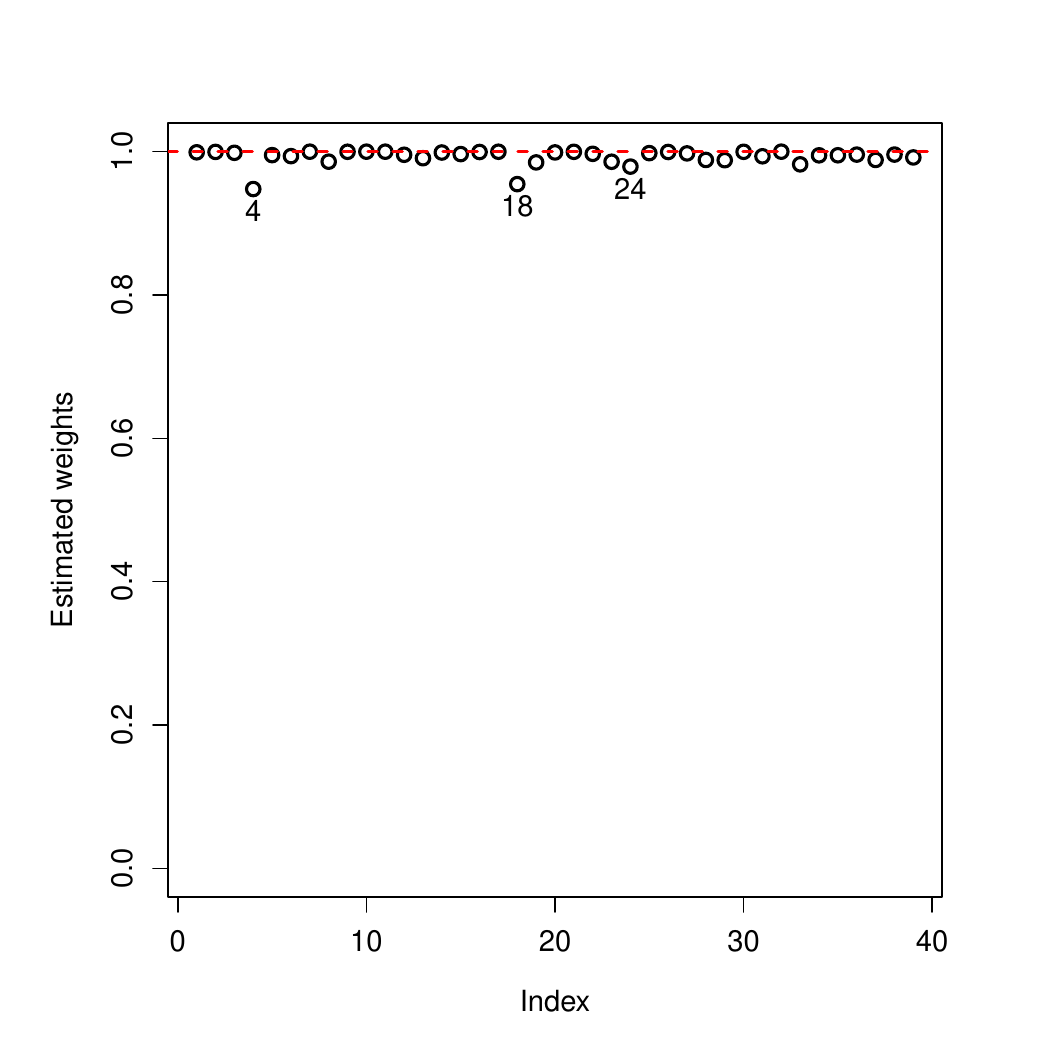}
  }
  \subfigure[$q = 0.96$]{
    \includegraphics[width = 0.28\linewidth]{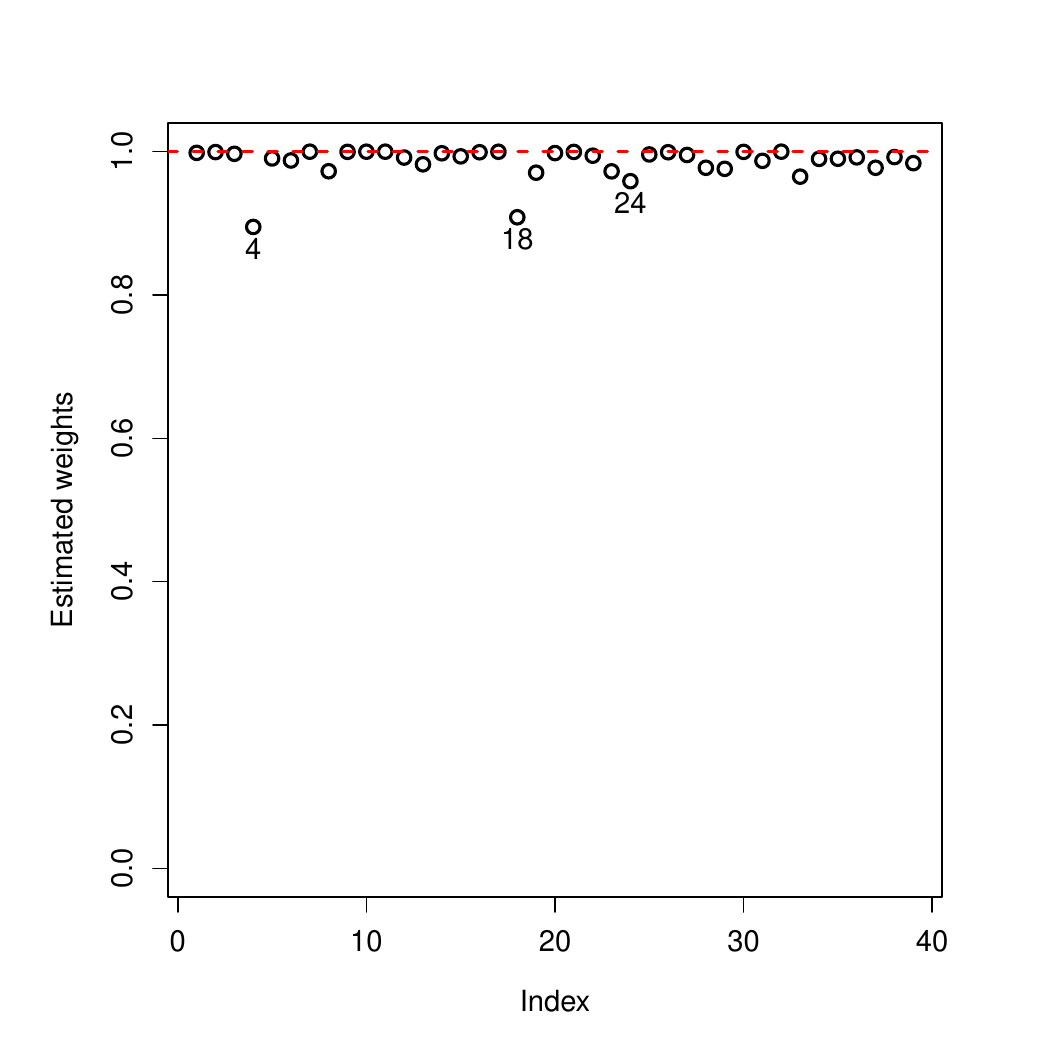}
  }
  \subfigure[$q = 0.94$]{
    \includegraphics[width = 0.28\linewidth]{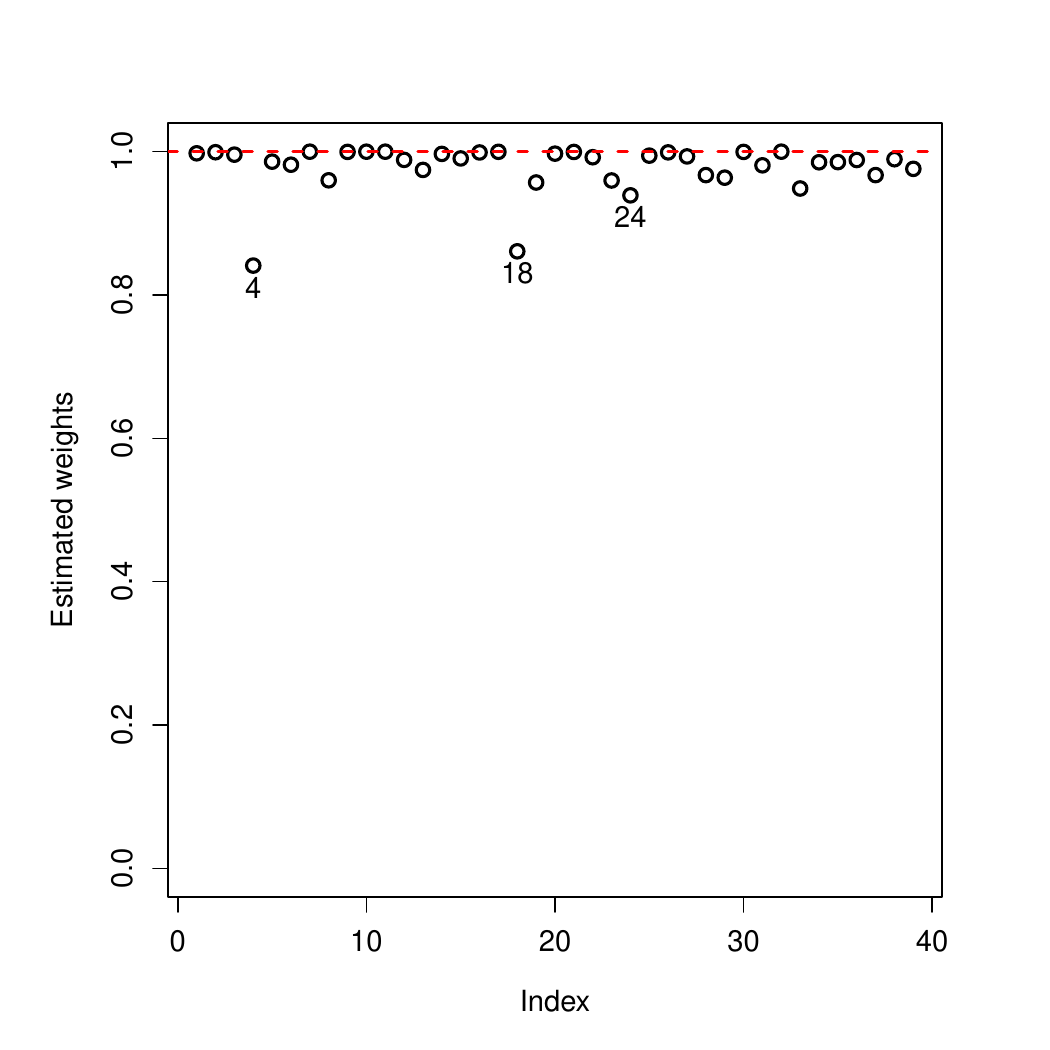}
  }
  \subfigure[$q = 0.92$]{
    \includegraphics[width = 0.28\linewidth]{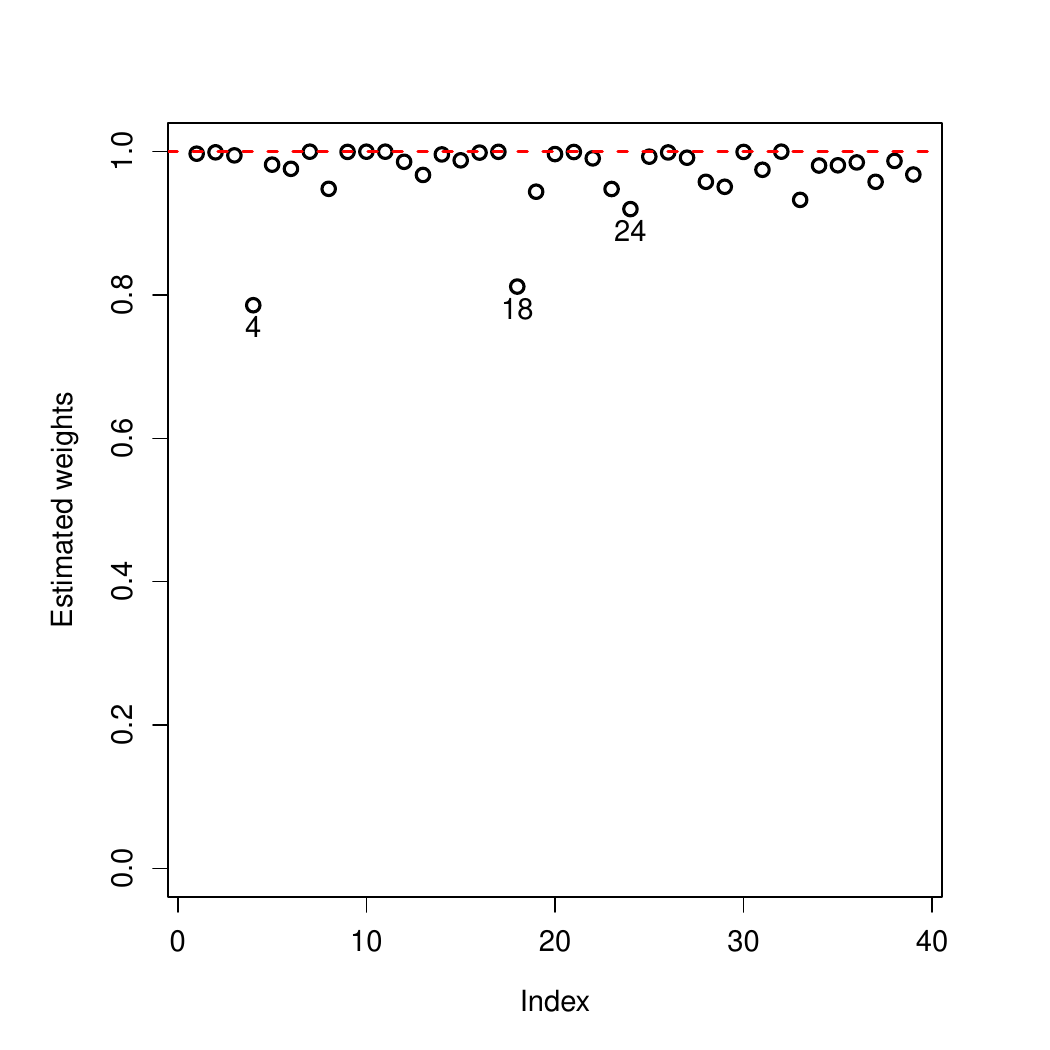}
  }
  \subfigure[$q = 0.90$]{
    \includegraphics[width = 0.28\linewidth]{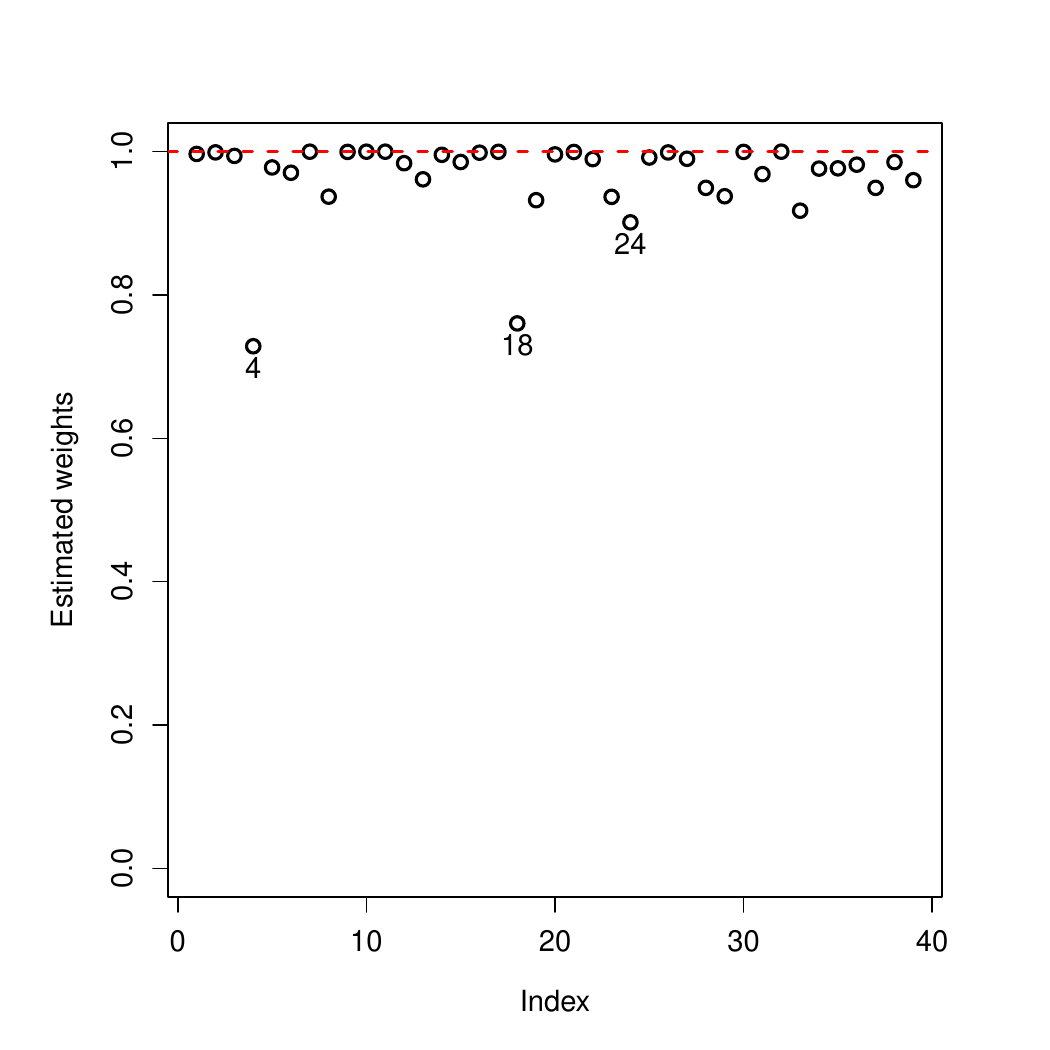}
  }
  \subfigure[$q = 0.88$]{
    \includegraphics[width = 0.28\linewidth]{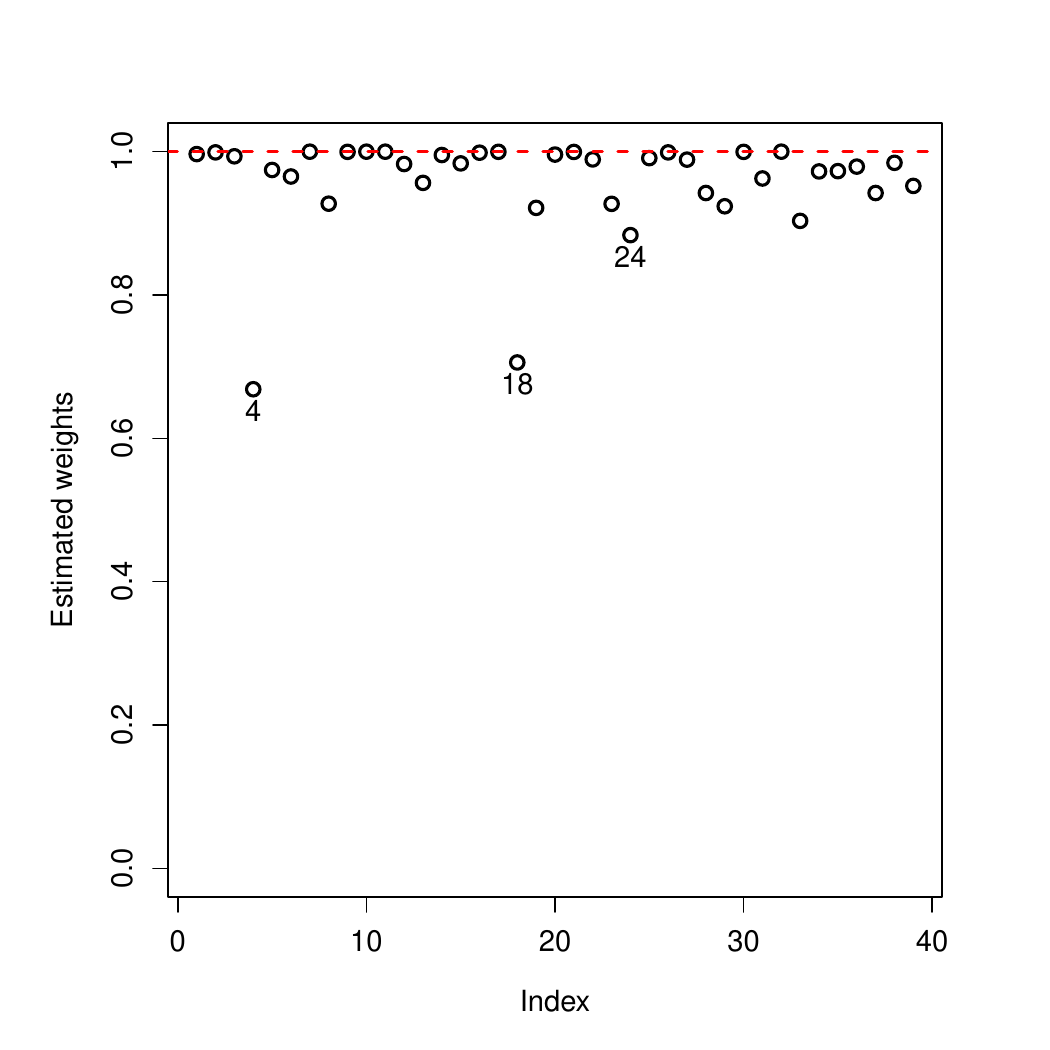}
  }
  \subfigure[$q = 0.86$]{
    \includegraphics[width = 0.28\linewidth]{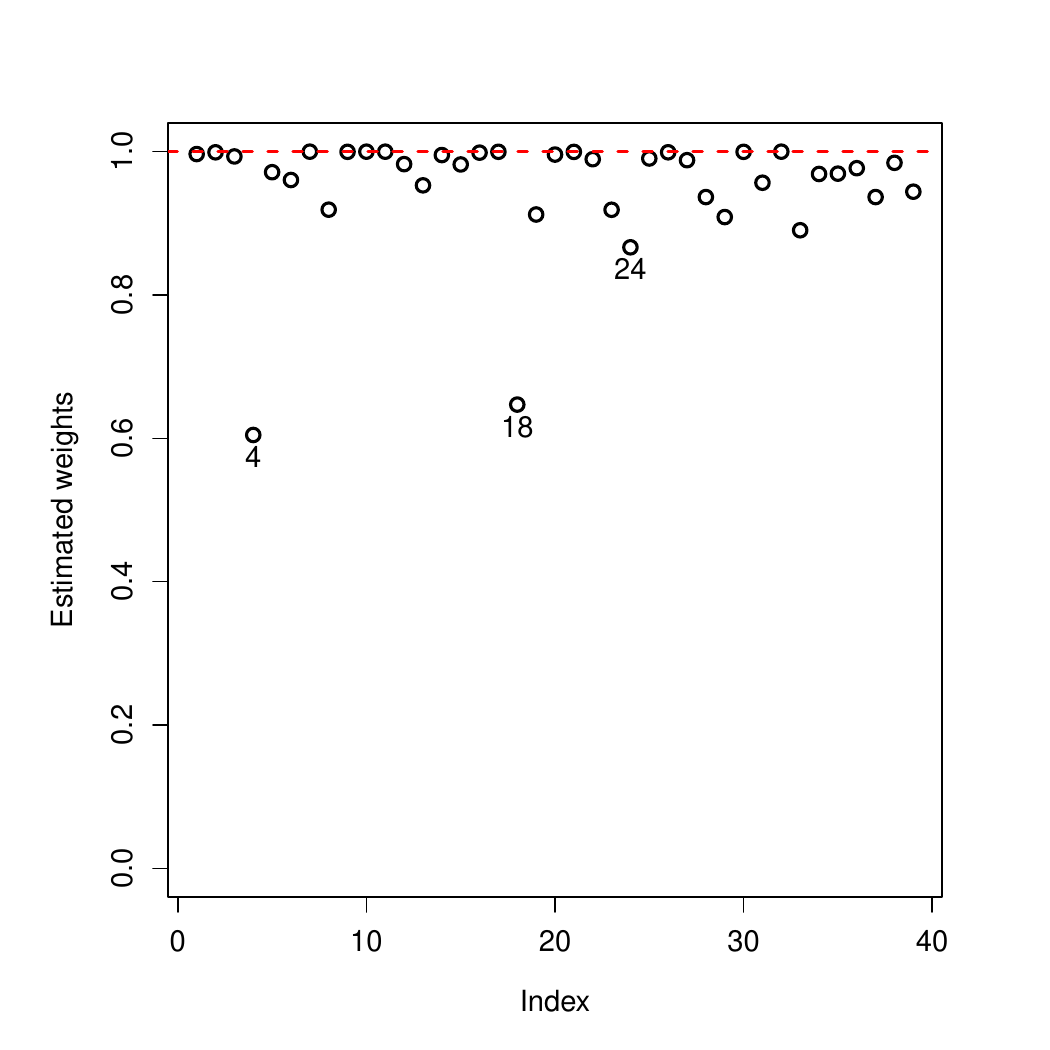}
  }
  \subfigure[$q = 0.84$]{
    \includegraphics[width = 0.28\linewidth]{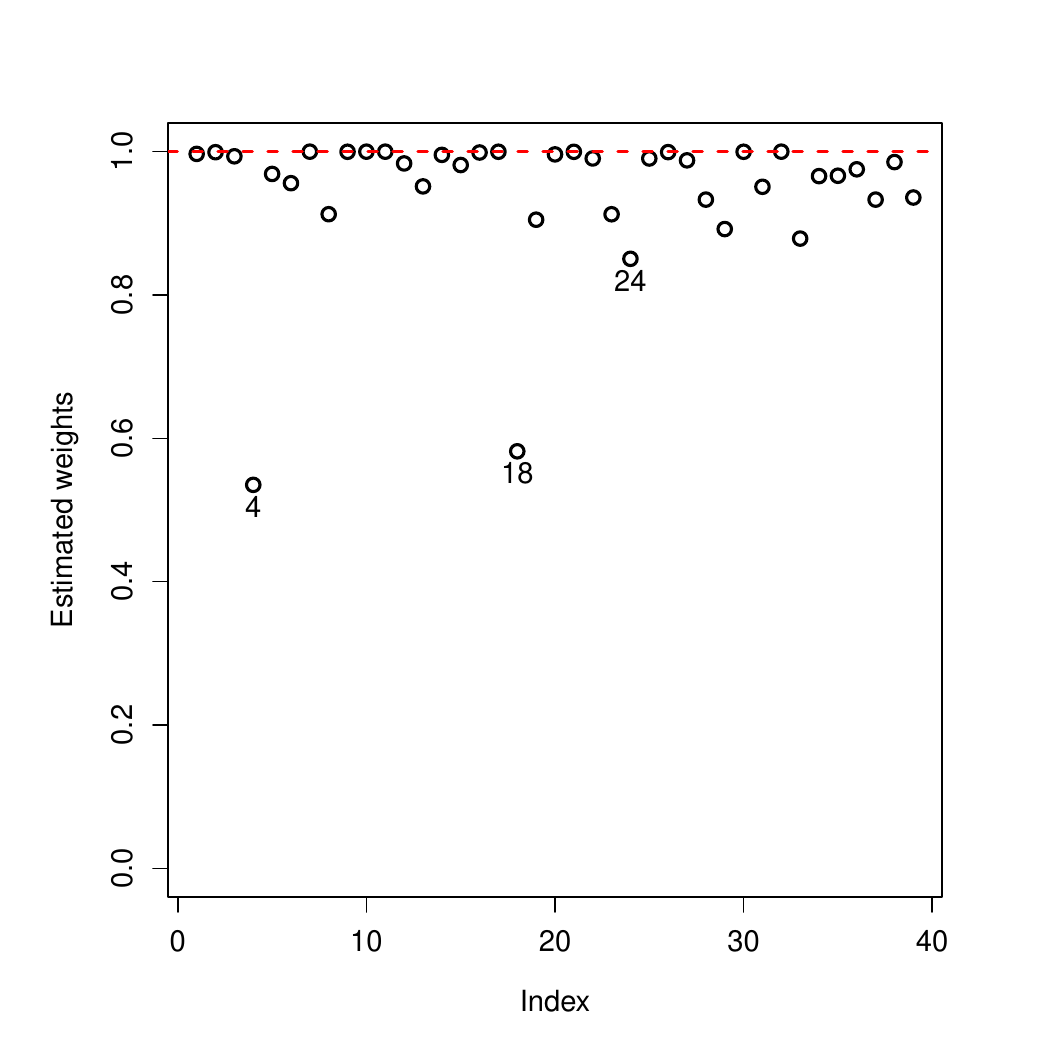}
  }
  \subfigure[$q = 0.82$]{
    \includegraphics[width = 0.28\linewidth]{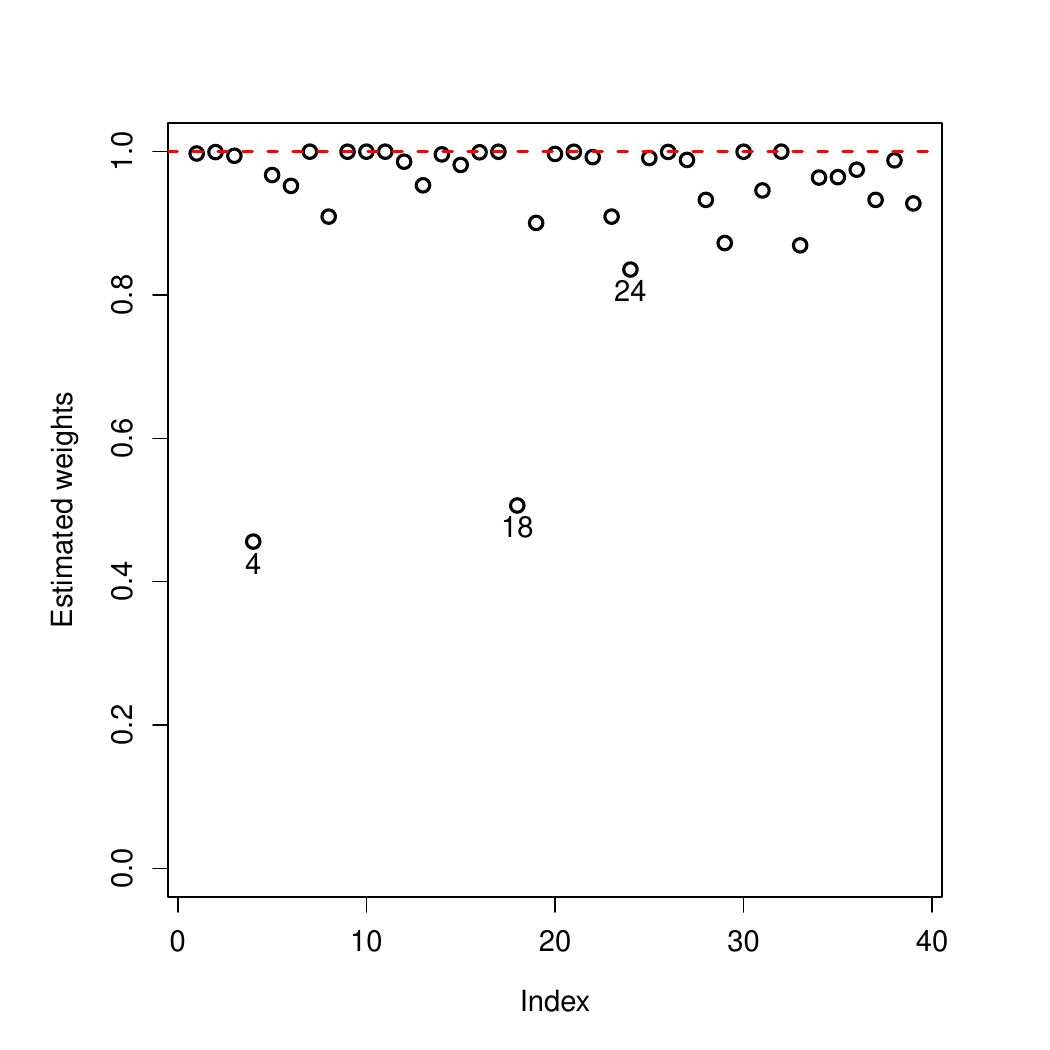}
  }
  \subfigure[$q = 0.78$]{
    \includegraphics[width = 0.28\linewidth]{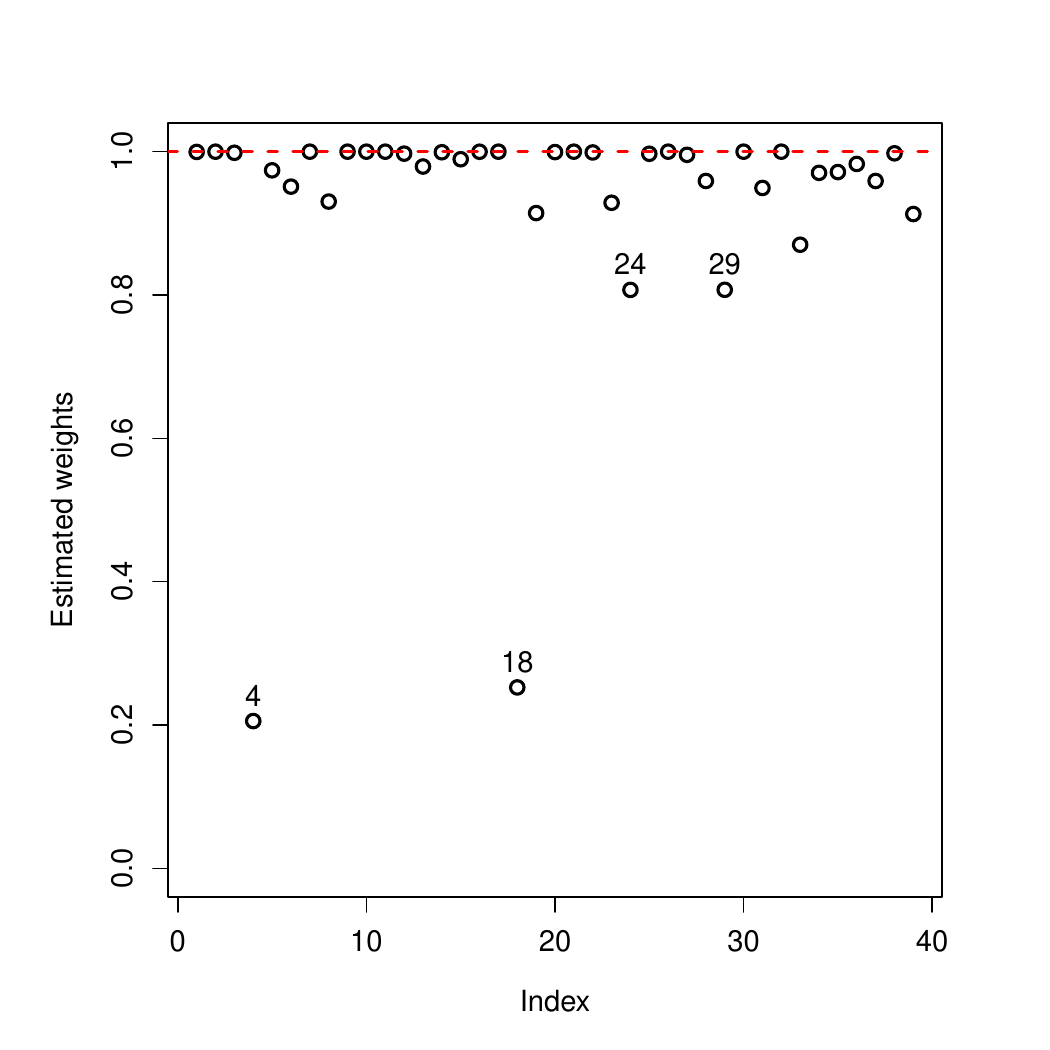}
  }
  \subfigure[$q = 0.76$]{
    \includegraphics[width = 0.28\linewidth]{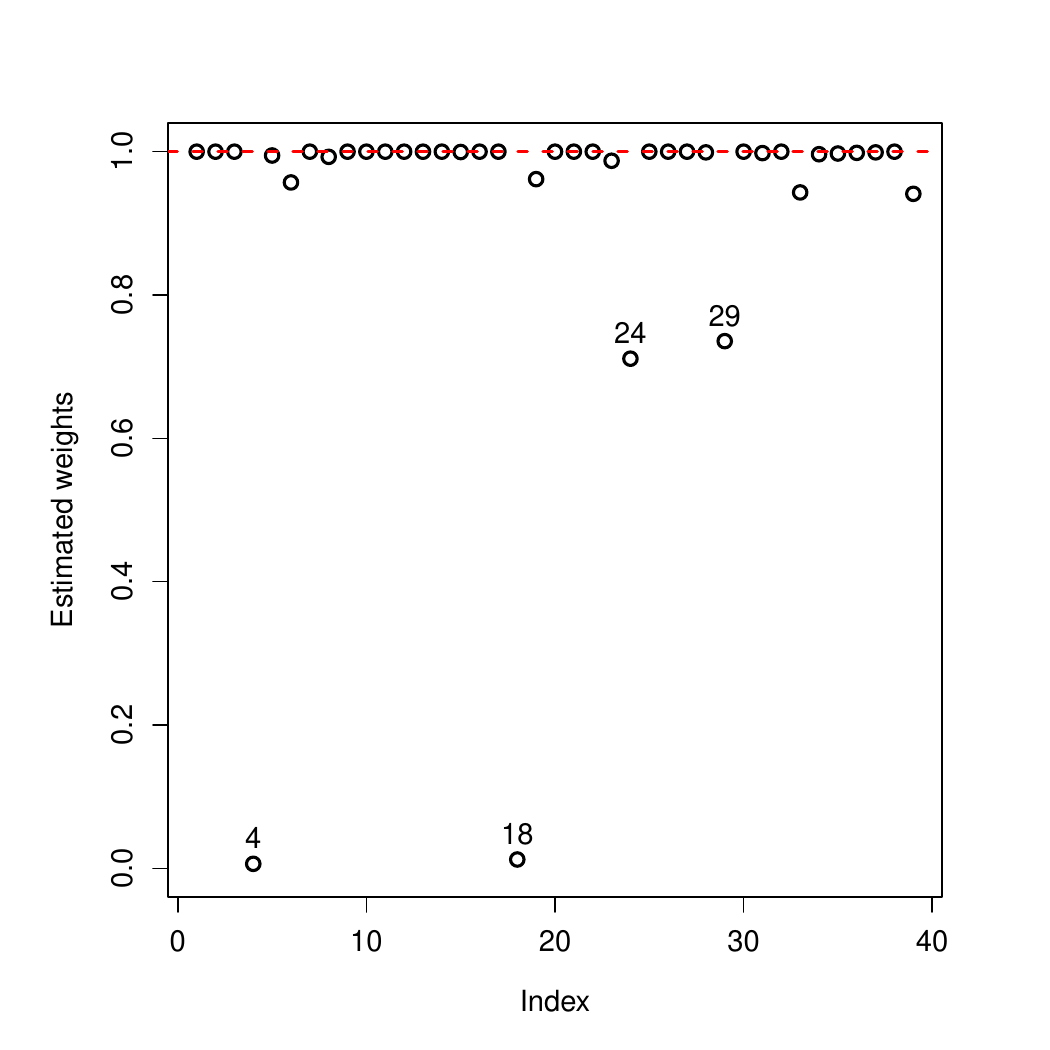}
  }
  \subfigure[$q = 0.74$]{
    \includegraphics[width = 0.28\linewidth]{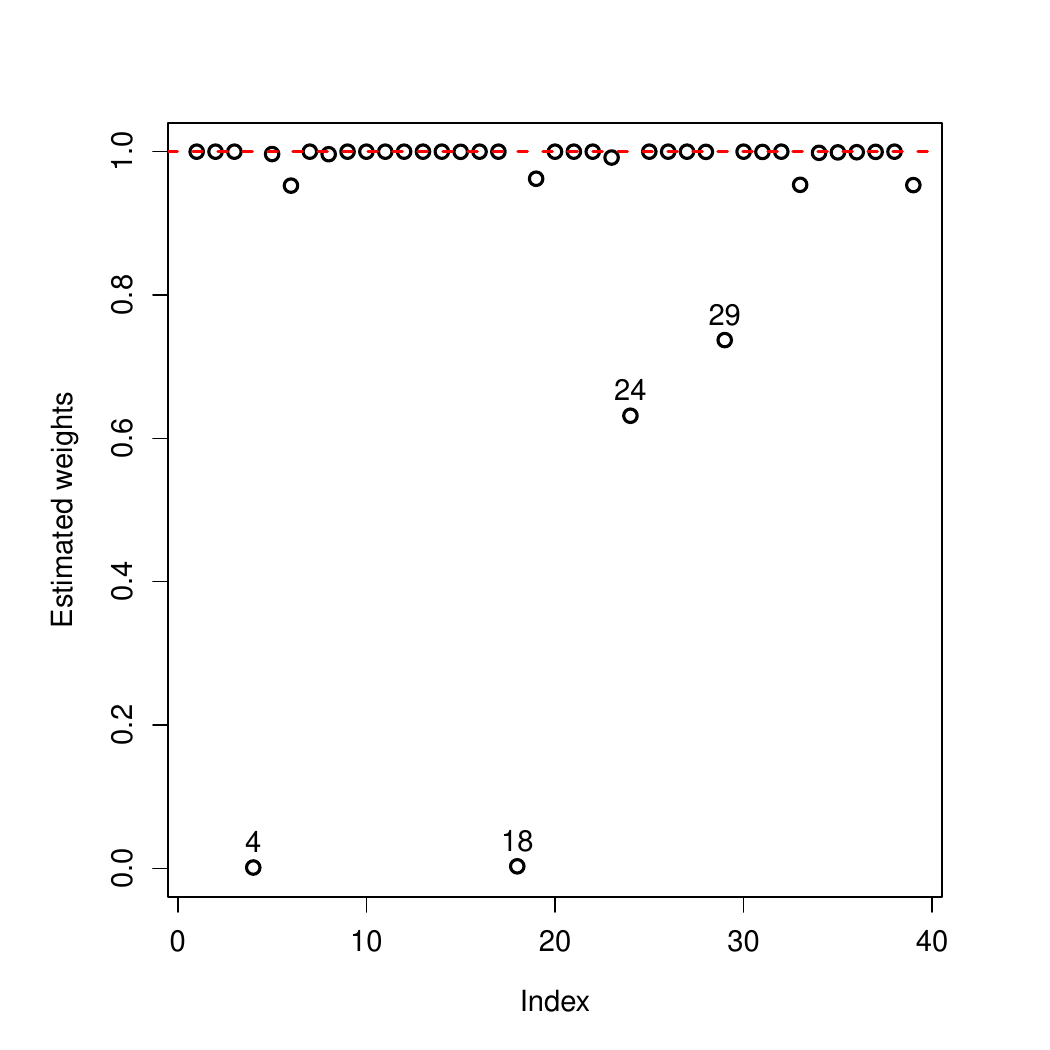}
  }
  \caption{Skin vaso-constriction data: estimated weights from the logistic
  model fitted using maximum L$q$-likelihood for several values of $q$.}\label{fig:wts}
\end{figure}

\newpage

\begin{figure}[!ht]
  \vskip -1em
  \centering
  \subfigure[$q = 0.98$]{
    \includegraphics[width = 0.28\linewidth]{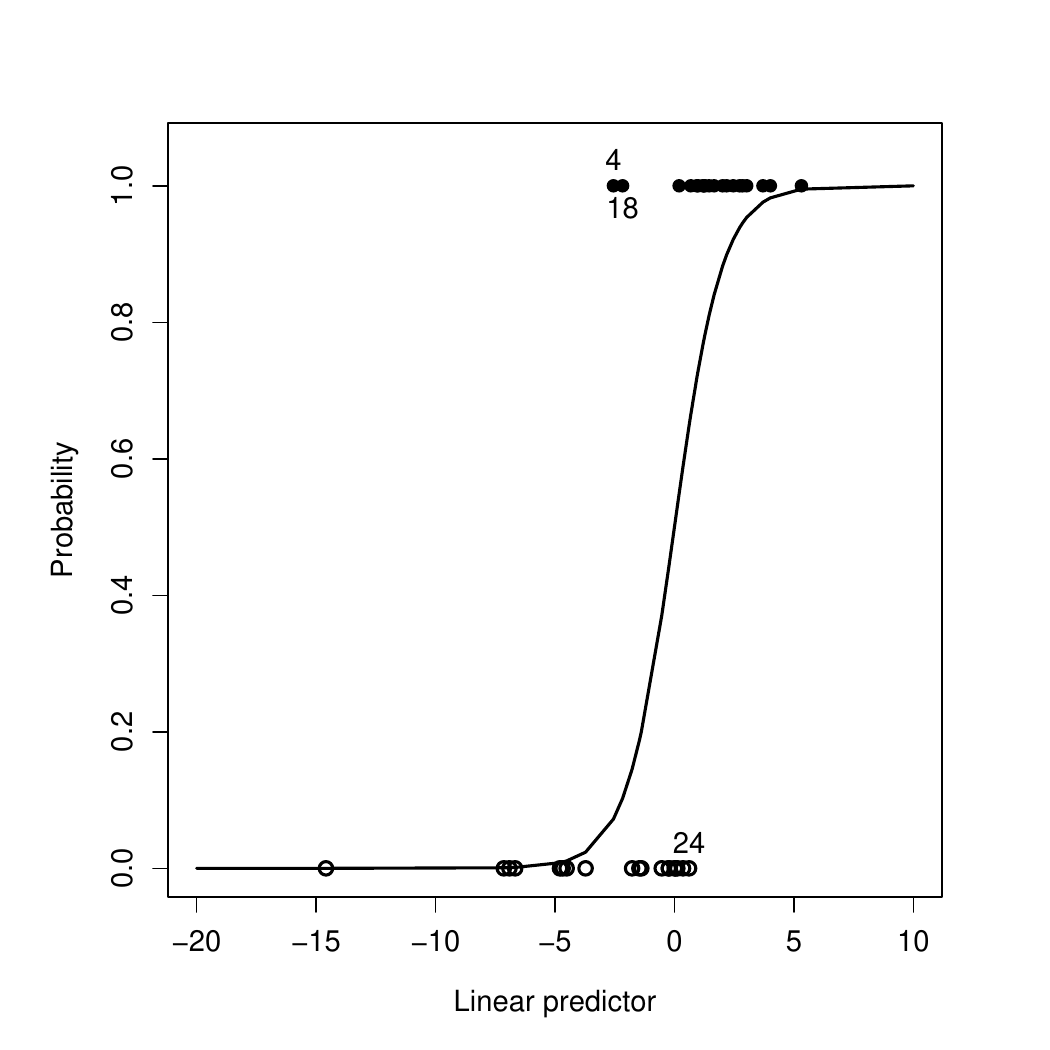}
  }
  \subfigure[$q = 0.96$]{
    \includegraphics[width = 0.28\linewidth]{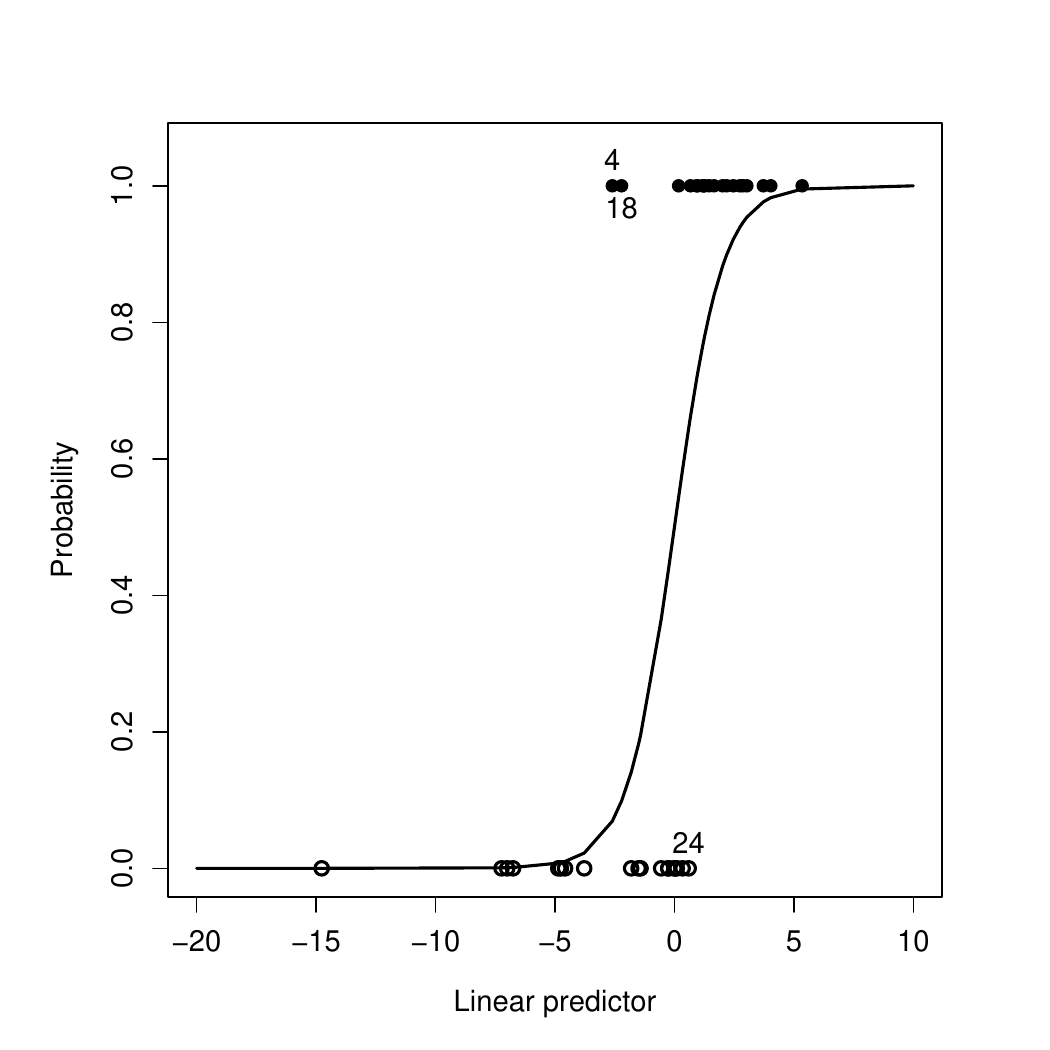}
  }
  \subfigure[$q = 0.94$]{
    \includegraphics[width = 0.28\linewidth]{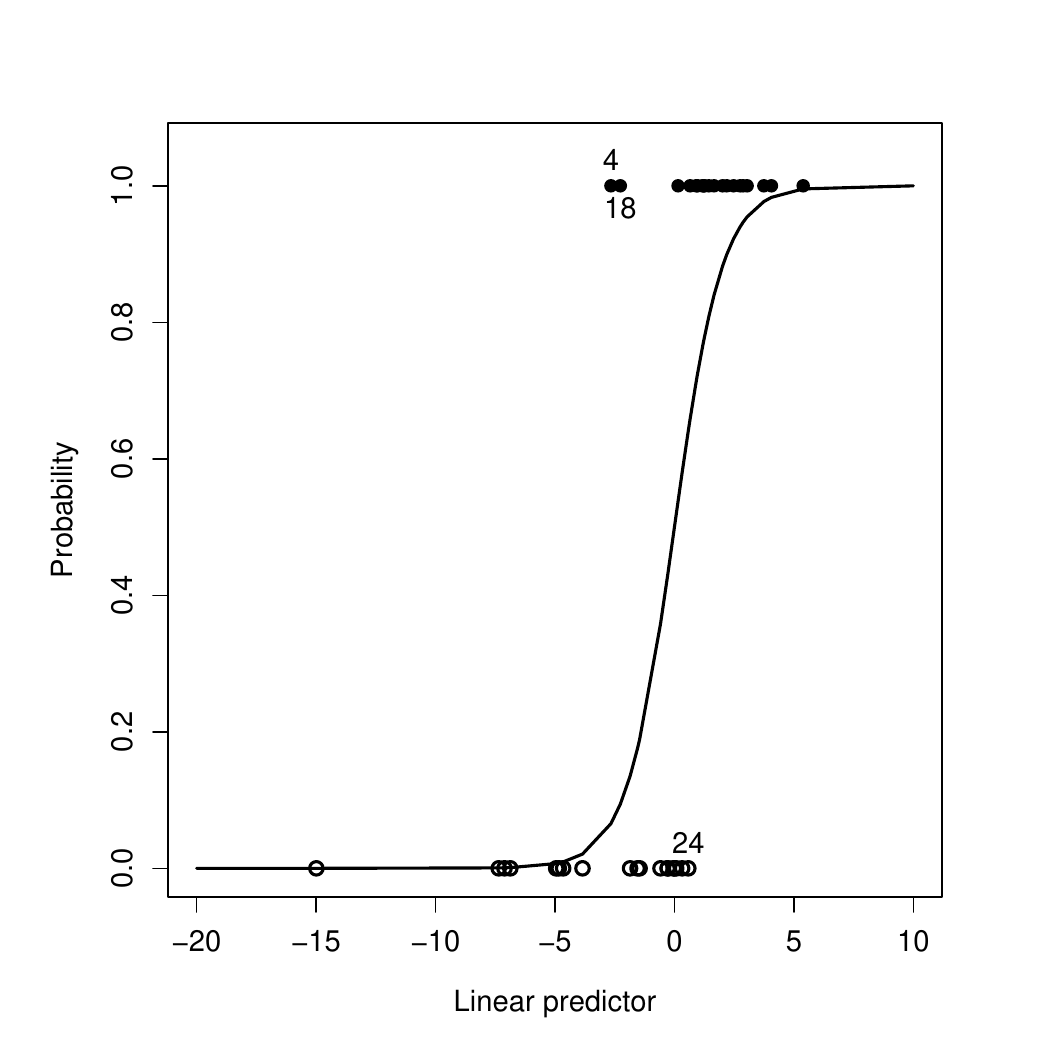}
  }
  \subfigure[$q = 0.92$]{
    \includegraphics[width = 0.28\linewidth]{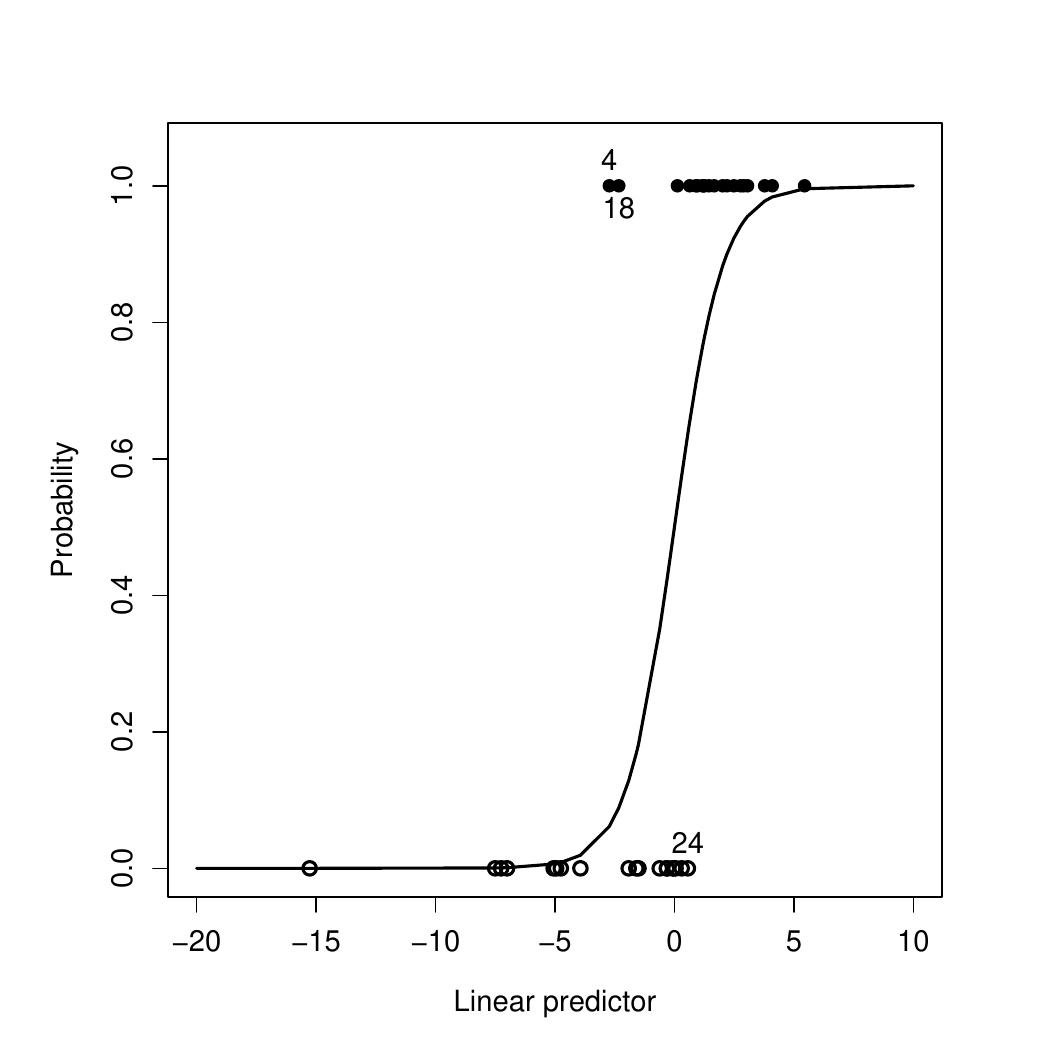}
  }
  \subfigure[$q = 0.90$]{
    \includegraphics[width = 0.28\linewidth]{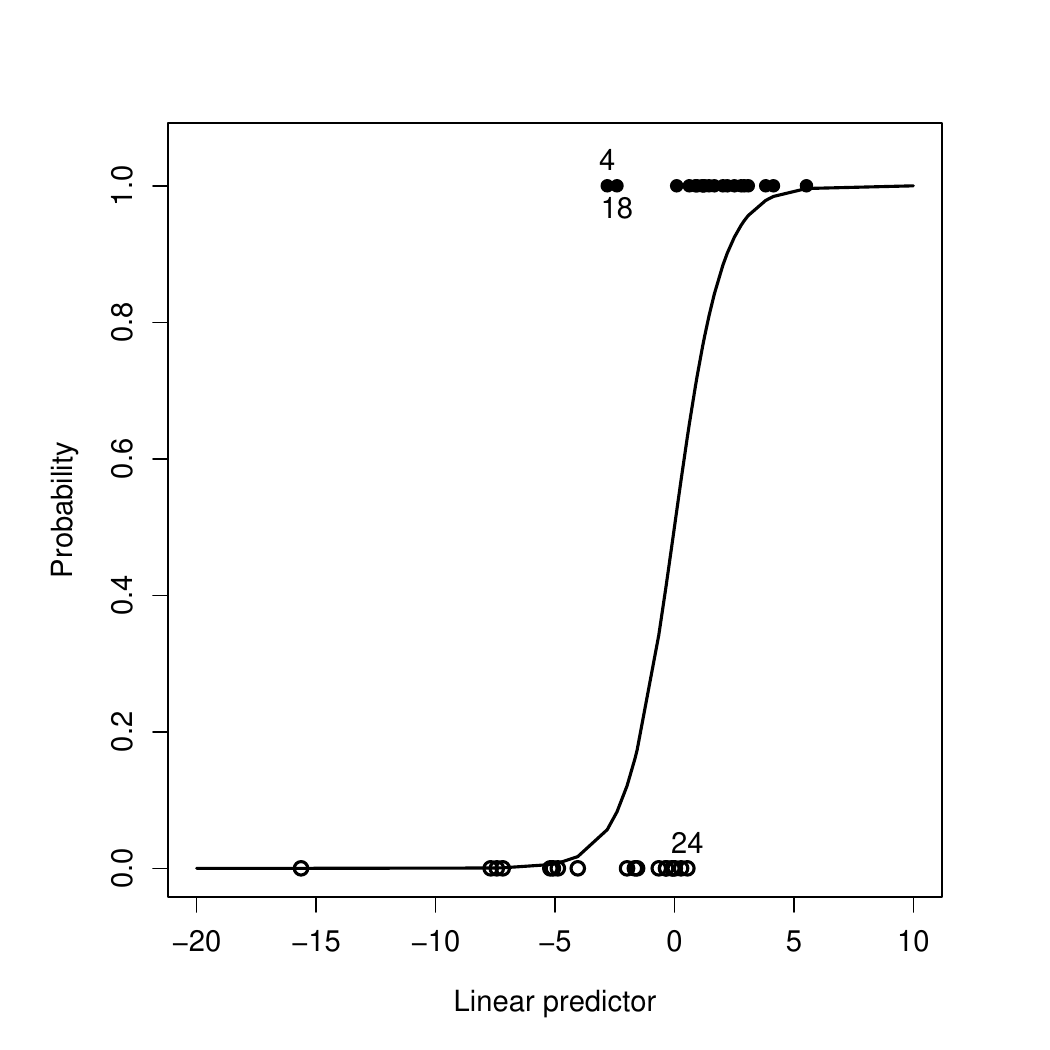}
  }
  \subfigure[$q = 0.88$]{
    \includegraphics[width = 0.28\linewidth]{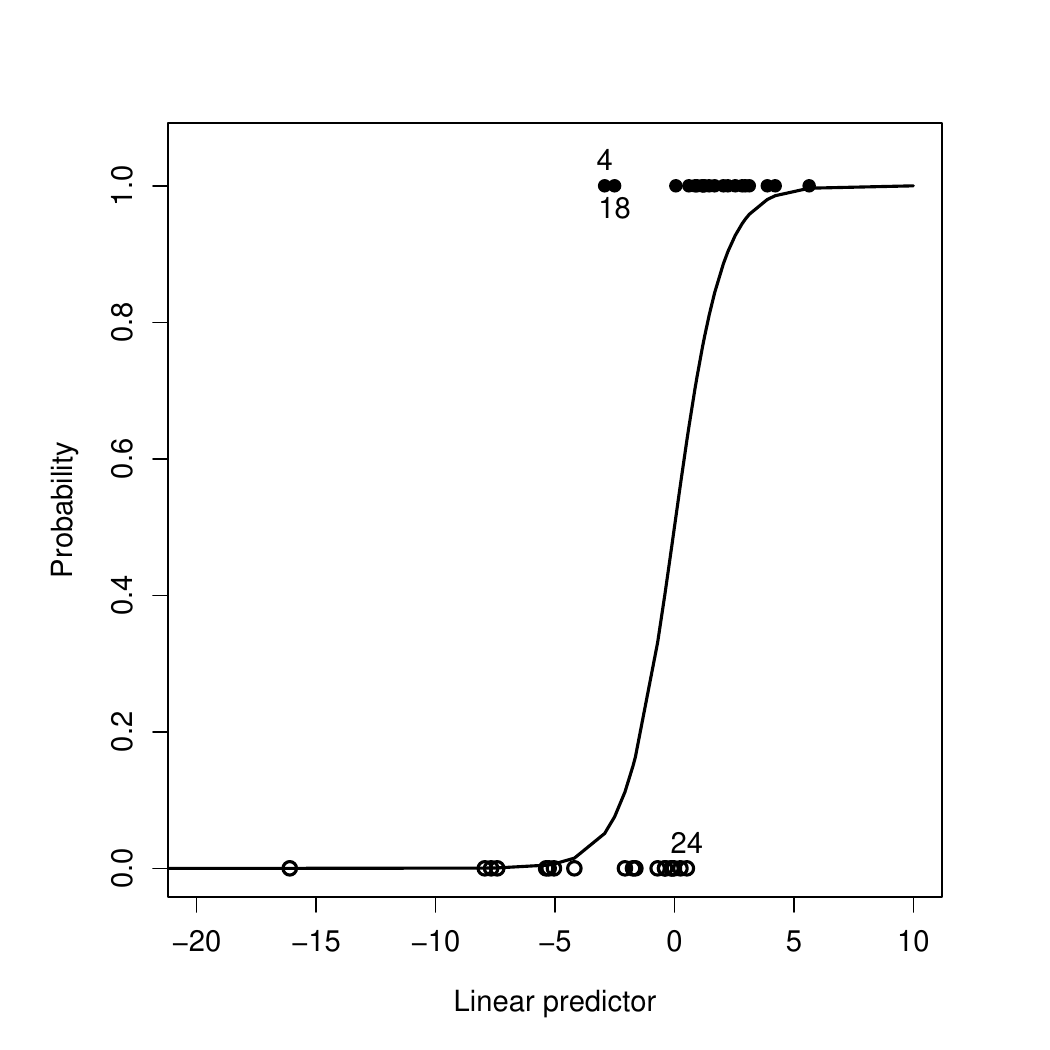}
  }
  \subfigure[$q = 0.86$]{
    \includegraphics[width = 0.28\linewidth]{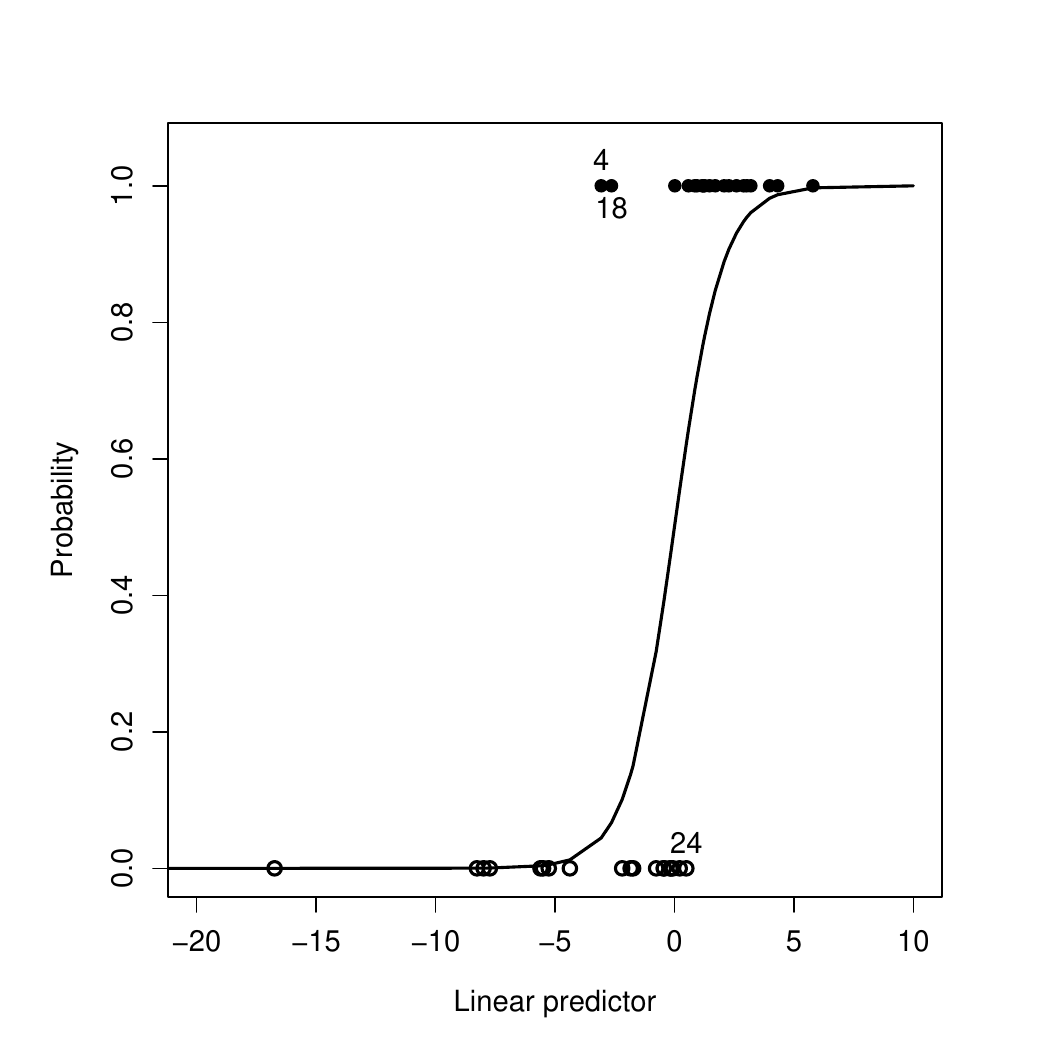}
  }
  \subfigure[$q = 0.84$]{
    \includegraphics[width = 0.28\linewidth]{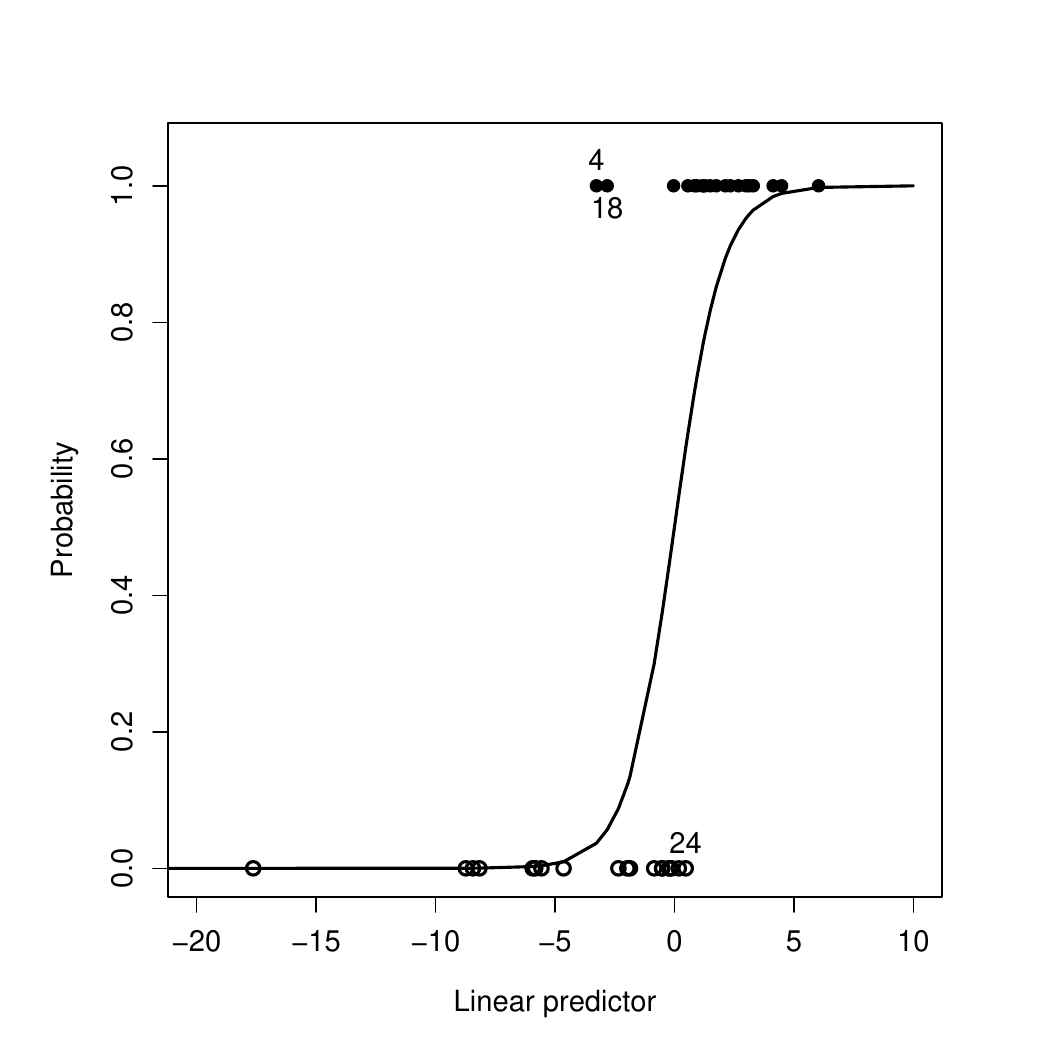}
  }
  \subfigure[$q = 0.82$]{
    \includegraphics[width = 0.28\linewidth]{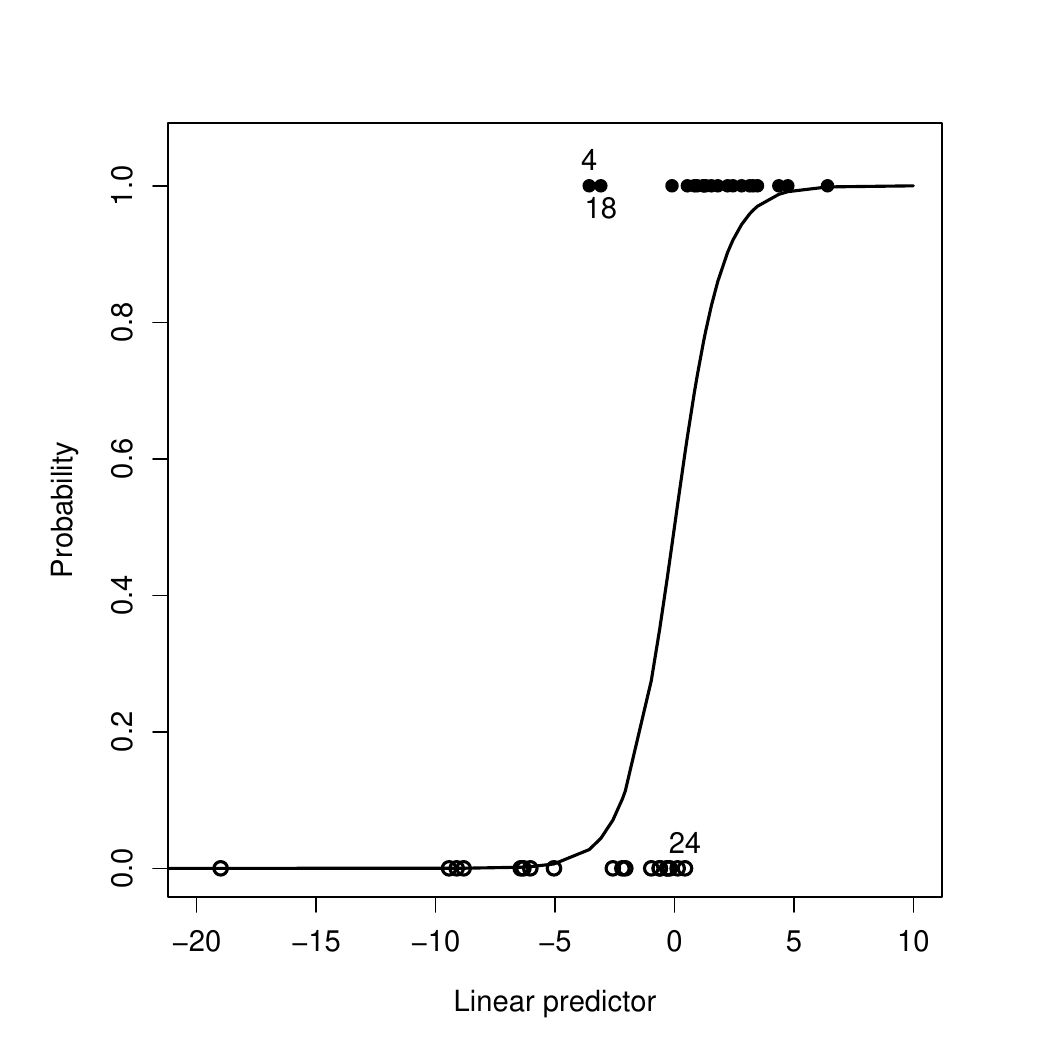}
  }
  \subfigure[$q = 0.78$]{
    \includegraphics[width = 0.28\linewidth]{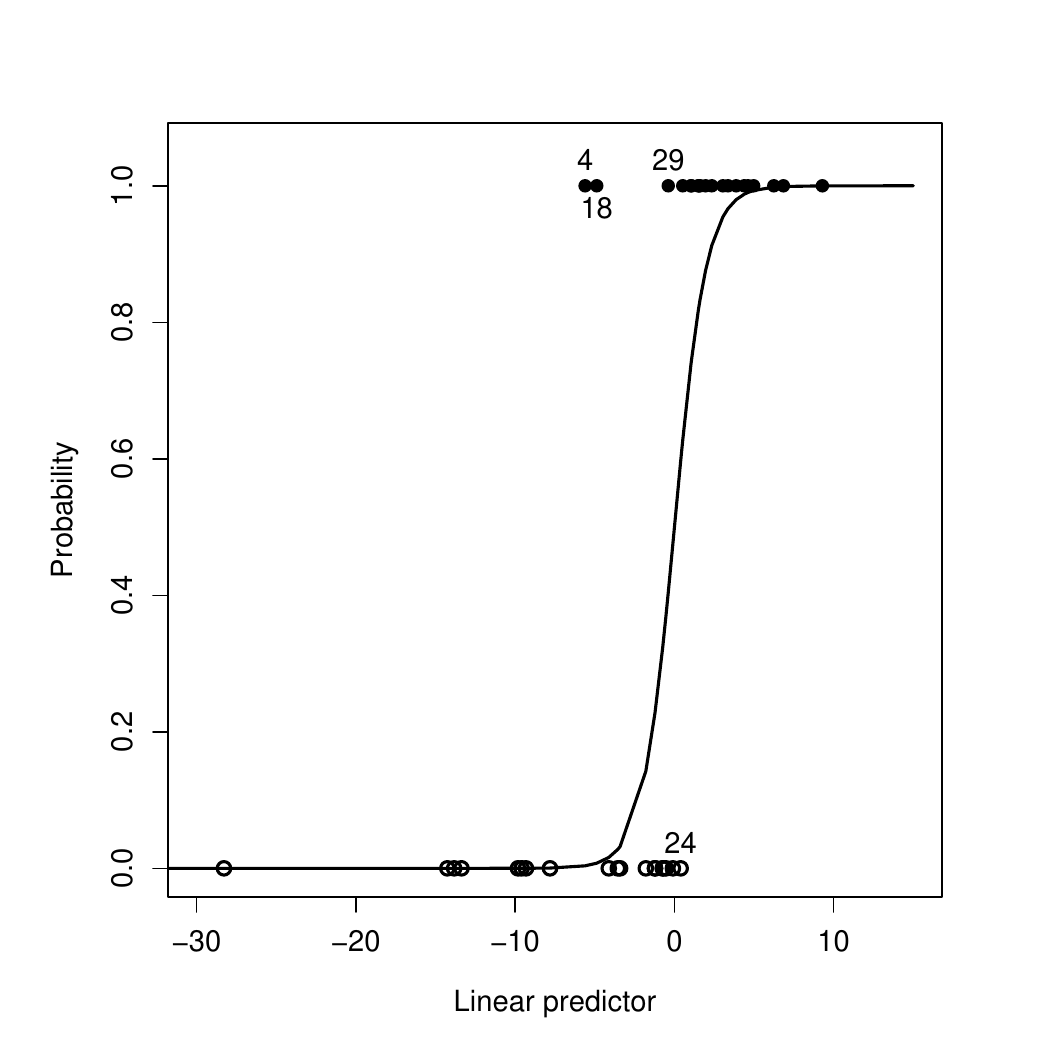}
  }
  \subfigure[$q = 0.76$]{
    \includegraphics[width = 0.28\linewidth]{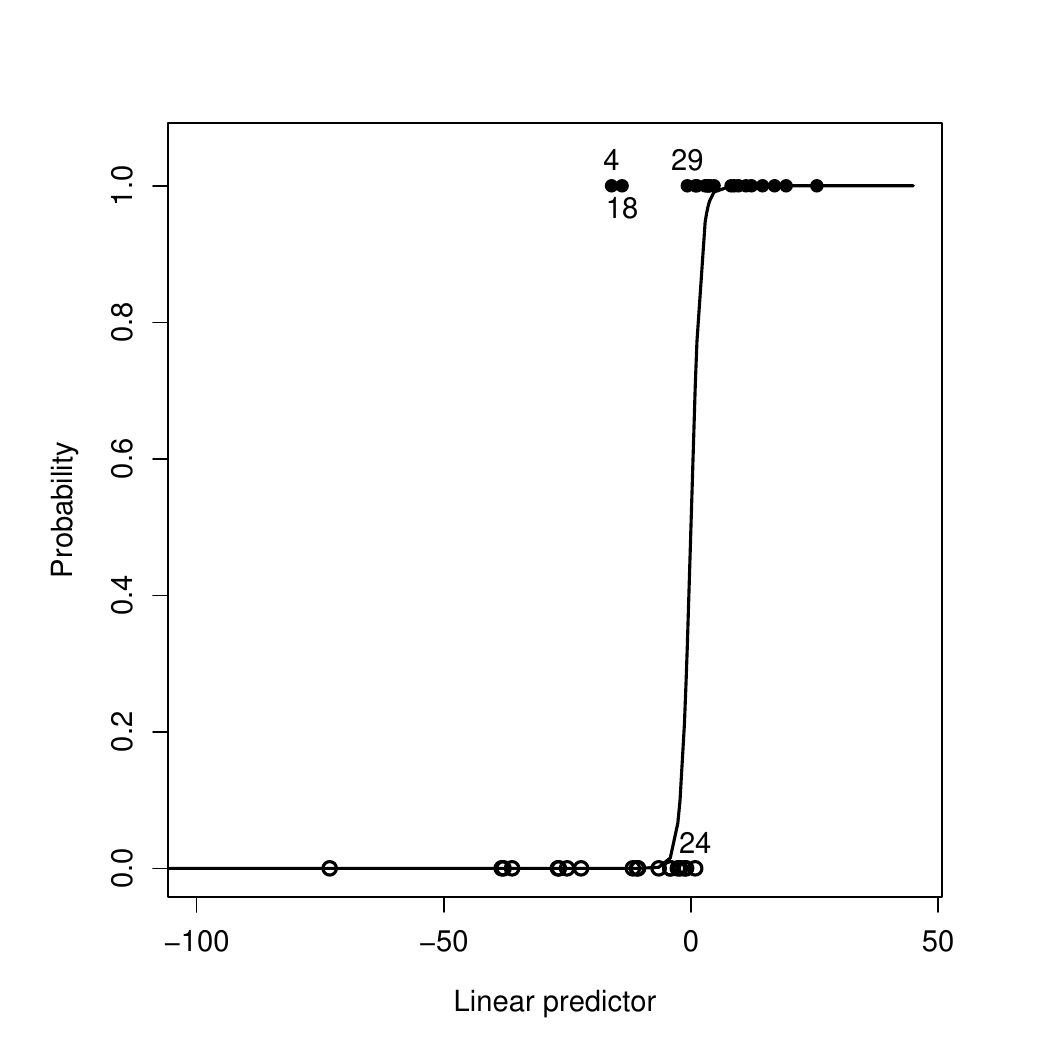}
  }
  \subfigure[$q = 0.74$]{
    \includegraphics[width = 0.28\linewidth]{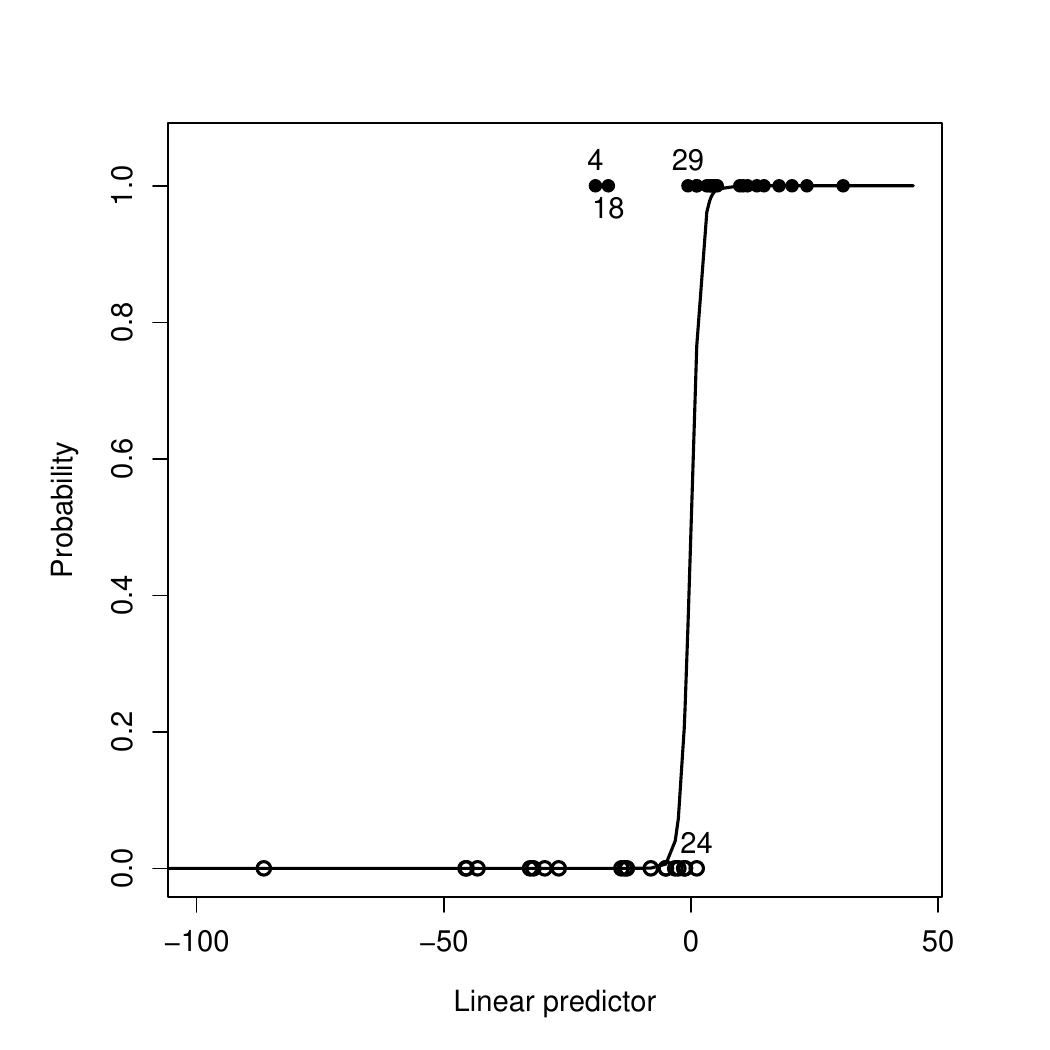}
  }
  \caption{Skin vaso-constriction data: observations and the estimated probabilities
  using maximum L$q$-likelihood for several values of $q$.}\label{fig:prob}
\end{figure}

\newpage

\bigskip

\end{document}